\documentclass[pra,print,superscriptaddress,twocolumn]{revtex4}
\usepackage{amssymb}
\usepackage{amsmath}
\usepackage{epsfig}
\usepackage{color}
\usepackage{graphics, graphicx}
\usepackage{bbold}
\usepackage{psfrag}
\usepackage{mathcomp}
\usepackage{subfigure}
\usepackage{verbatim}
\usepackage{color}
\usepackage[colorlinks,citecolor=blue]{hyperref}

\begin{document}

\title{Atomic self-organization emerging from tunable quadrature coupling }
\author{Jingtao Fan}
\affiliation{State Key Laboratory of Quantum Optics and Quantum Optics Devices, Institute
of Laser Spectroscopy, Shanxi University, Taiyuan 030006, China}
\affiliation{Collaborative Innovation Center of Extreme Optics, Shanxi University,
Taiyuan 030006, China}
\author{Gang Chen}
\thanks{chengang971@163.com}
\affiliation{State Key Laboratory of Quantum Optics and Quantum Optics Devices, Institute
of Laser Spectroscopy, Shanxi University, Taiyuan 030006, China}
\affiliation{Collaborative Innovation Center of Extreme Optics, Shanxi University,
Taiyuan 030006, China}
\affiliation{Collaborative Innovation Center of Light Manipulations and Applications,
Shandong Normal University, Jinan 250358, China}
\author{Suotang Jia}
\affiliation{State Key Laboratory of Quantum Optics and Quantum Optics Devices, Institute
of Laser Spectroscopy, Shanxi University, Taiyuan 030006, China}
\affiliation{Collaborative Innovation Center of Extreme Optics, Shanxi University,
Taiyuan 030006, China}

\begin{abstract}
The recent experimental observation of dissipation-induced structural
instability provides new opportunities for exploring the competition
mechanism between stationary and nonstationary dynamics [Science \textbf{366}%
, 1496 (2019)]. In that study, two orthogonal quadratures of cavity field
are coupled to two different Zeeman states of a spinor Bose-Einstein
condensate (BEC). Here we propose a scheme to couple two density-wave
degrees of freedom of a BEC to two quadratures of the cavity field.
Different from previous studies, the light-matter quadratures coupling in
our model is endowed with a tunable coupling angle. Apart from the uniform
and self-organized phases, we unravel a dynamically unstable state induced
by the cavity dissipation. Interestingly, the dissipation defines a
particular coupling angle, across which the instabilities disappear.
Moreover, at this critical coupling angle, one of the two atomic density
waves can be independently excited without affecting one another. It is also
found that our system can be mapped into a reduced three-level model under
the commonly used low-excitation-mode approximation. However, the
effectiveness of this approximation is shown to be broken by the dissipative
nature for some special system parameters, hinting that the
low-excitation-mode approximation is insufficient in capturing some
dissipation-sensitive physics. Our work enriches the quantum simulation
toolbox in the cavity-quantum-electrodynamics system and broadens the
frontiers of light-matter interaction.
\end{abstract}

\maketitle

\section{Introduction}

Dissipative quantum many-body system lies at the heart of diverse branches
of physics such as statistical mechanics, condensed matter physics, and
quantum optics \cite{Book2}. Compared to its equilibrium analog, a system
exposed to dissipation is even harder to be understood due to the somewhat
uncontrolled environment couplings. Fortunately, with the rapid improvement
of both experimental and theoretical techniques, lots of exciting progress
in this realm have been made \cite%
{NATJK06,SCIN08,NATA09,PRLN10,PNASF13,PRLG13,PNASJ15,NPT18,SAT17,SAS19,PRXJ20,PRLS12,PRLB12,NPS11,PRLS10,PRAD11,PRLL13,PRLM18,PRLH20}%
. It has been shown that the interplay between coherent and dissipative
dynamics can lead to a vast kinds of novel phenomena. Examples include
nonequilibrium transition \cite%
{NATJK06,SCIN08,NATA09,PRLN10,PNASF13,PRLG13,PNASJ15,NPT18,SAT17,SAS19,PRXJ20}%
, interaction-mediated laser cooling \cite{PRLS12,PRLB12}, topological
effects \cite{NPS11}, dynamical new universality classes \cite%
{PRLS10,PRAD11,PRLL13}, and multistability of quantum spins \cite%
{PRLM18,PRLH20}. Among various realizations of the dissipative system, the
coherently driven atomic gases inside optical cavities emerge as a uniquely
promising route \cite%
{RMPR13,NATK10,NATJ17,SCIJ17,SCIR12,PRLJ15,NATR16,PRLS15,PRAS16,PRLR15,PRAC16,PRBT17,PRLZ16,PRLC16,PRLK17,Bikash14,PRLJP15,PRLF17,EPJD08,NPS09,dicketheory1,dicketheory2,dicketheory3,dicketheory4,dicketheory5,cavfermion1,EPJDD08,Fan14,Feng18,Guan19,PRLSP20}%
. Photons leaking from the cavity not only provide a
convenient way to probe the atomic state, but also\ open a controlled
channel for the collective dissipative dynamics \cite%
{PRAP03,PRLI07,PRLI09,NCR15,SCIN19,NJPK20,PRLR18,PNASL18}. Moreover, the scattered cavity photons feed back on the atomic degrees of
freedom and effectively impose a dynamic potential \cite%
{SCIR12,PRLJ15,NATR16,PRLS15,PRAS16}, which favors a unitary evolution of
atoms. The competition between the coherent and dissipative processes in
this composite system are fairly responsible for interesting nonequilibrium
collective dynamics and exotic steady states.

Recently, plenty of noticeable effects induced by the driven-dissipative
nature of the atom-cavity system have been uncovered both experimentally
\cite%
{PNASL18,PRLR18,PRAY19,PRXV18,PRLR19,PRLY19,PRLP19,PRAA19,OPTIC19,NJPK20}
and theoretically \cite%
{Fan18,PRLFM18,PRLFM19,PRLK19,PRLF18,ARCC19,ARCC20,PRLS20}. The light-matter
interaction considered by these studies has been, however, mostly limited to
the coupling between an atomic density mode and a single quadrature of
cavity fields, which loses potential physics rooted in the cooperative
interplay among multiple light quadratures. Actually, the combined action of
the two orthogonal quadratures may have major impacts on spin systems \cite%
{PRAP12,PRLA14}. For example, it has been predicted that the simultaneous
coupling between quantum spins and the two orthogonal quadratures of a
radiation field can lead to anomalous multicritical points \cite{PRLM18}.
Along the same research direction, some judicious experiments impose this
type of coupling on two different Zeeman states of a spinor BEC \cite%
{PRLML18,SCIN19}, demonstrating that the competition between coherent and
dissipative processes can even trigger a structural instability \cite{SCIN19}%
. This progress further advances a series of relevant theoretical works \cite%
{DampT1,DampT2,DampT3}. Nevertheless, given that the quadrature operator of
light is characterized by a phase factor representing a rotation angle
(dubbed coupling angle) in the phase space \cite{Book01}, these researches
focus only on the orthogonal light-atom coupling case where the coupling
angle is frozen to $\pi /2$, leaving the interaction mechanism arising from
a more generic coupling angle largely unexplored. This encourages us to
raise the following fundamental questions: (i) what new physics may emerge
from the light-matter interaction if the involved quadratures of radiation
field can be tuned via the coupling angle? (ii) what is the role of
dissipation in such a system?

In this paper, we address these questions by studying a driven-dissipative
BEC-cavity system. We propose an experimental scheme, where two density-wave
degrees of freedom of the BEC are coupled to two quadratures of the cavity
field. In contrast to previous proposals, here the two quadratures of the
cavity field carry a coupling angle $\theta $, which, together with their
respective pump strengths, can be feasibly controlled in experiment.

Apart from the uniform and self-organized phases, we unravel a dynamically
unstable state induced by the cavity dissipation. By adiabatically
eliminating the cavity field, we show that the dissipation defines a
particular coupling angle $\theta _{c}$, across which the instabilities
completely disappear. More importantly, when the coupling angle equals $%
\theta _{c}$, one of the two density modes can be independently excited
without affecting one another. Going beyond the adiabatic elimination, the
normal phase becomes unstable. The instabilities coming from the
nonadiabaticity, however, turn out to be negligible for typical parameters
in the current experiments. It is also found that our system can be mapped
into a reduced three-level model under the commonly used low-excitation-mode
approximation. However, we show the dissipative nature could break the
effectiveness of the three-level model for some parameters, hinting that the
low-excitation-mode approximation may be questionable in capturing some
dissipation-sensitive physics.

The work is organized as follows. In Sec.~\ref{sec:system}, we describe the
proposed system configuration and present the Hamiltonian. In Sec.~\ref%
{sec:approach}, we present the mean-field approach used in calculating the
phase diagrams. In Sec.~\ref{sec:clophase}, we calculate the phase diagrams
for the closed system. In Sec.~\ref{sec:stability}, we carry out a stability
analysis and characterize the effects of dissipation on the system. In Sec.~%
\ref{sec:dissphase}, we show the steady-state phase diagrams for the
driven-dissipative system. In Sec.~\ref{sec:nonadiabaticity}, we go beyond
the adiabatic elimination by including the dynamics of the cavity
fluctuations. In Sec.~\ref{sec:threemode}, we map the system into a reduced
three-level model by the three-mode approximation. We discuss the
experimental implementation in Sec.~\ref{sec:experiment}, and summarize in
Sec.~\ref{sec:conclusion}.
\begin{figure}[tp]
\includegraphics[width=8cm]{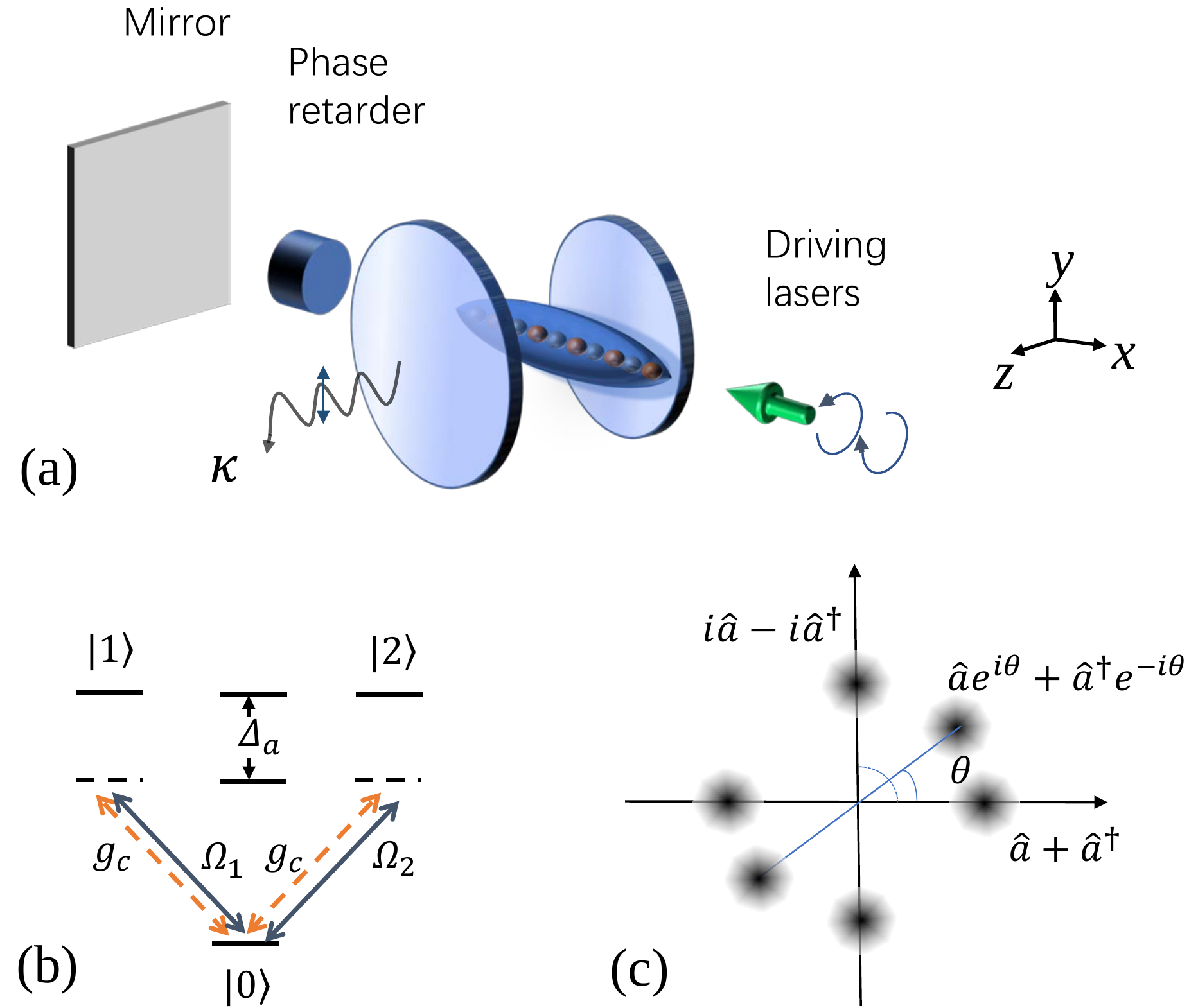}
\caption{(a) Schematic illustration of the considered setup. A quasi-1D BEC
is illuminated by a pair of orthogonally-polarized lasers that is
backreflected by a mirror. The phase retarder sitting in between the mirror
and the BEC produces polarization-dependent optical lengths for traversed
laser lights, and thereby imparts an additional phase shift between the two
backforward propagating light fields. (b) The atomic level scheme. The atoms
are simultaneously driven by the cavity field (red dashed arrows) and the
counterpropagating lasers (blue solid arrows).(c) Sketch of the field
distribution in the phase space responsible for different cavity
quadratures. }
\label{setup}
\end{figure}

\section{System}

\label{sec:system}

As illustrated in Fig.~\ref{setup}(a), we consider a BEC prepared inside an
optical cavity and driven by a pair of orthogonally-polarized lasers. The
BEC is assumed to be a cigar shape (with length $L$) elongated along the $x$
direction, which we take as the quantization axis. The two driving lasers,
which are frequency degenerate but with independently tunable phases and
amplitudes, copropagate along the $x$ direction, forming a generic
elliptically-polarized single beam before impinging on the atoms. After
propagating through the BEC, this laser beam is then backreflected from a
mirror, and traverses the BEC a second time. A polarization-sensitive phase
retarder is placed in between the mirror and the BEC, imparting an
additional phase shift between the two orthogonally-polarized backforward
propagating fields. The incident lasers with the same polarizations couple
the electronic ground state $\left\vert 0\right\rangle $ of the atoms to two
excited states $\left\vert 1\right\rangle $ and $\left\vert 2\right\rangle $
with Rabi frequencies $\Omega _{1}$ and $\Omega _{2}$, respectively. The
optical cavity, whose main axis is arranged perpendicular to the long axis
of the BEC, singles out a specific quantization mode and typically enhances
its interaction with the atoms. The selected cavity mode simultaneously
mediates the transitions $\left\vert 0\right\rangle \longleftrightarrow
\left\vert 1\right\rangle $ and $\left\vert 0\right\rangle
\longleftrightarrow \left\vert 2\right\rangle $ with coupling strength $%
g_{c} $ [see Fig.~\ref{setup}(b)]. The cavity frequency $\omega _{c}$ is
closed to that of the driving lasers $\omega _{p}$, both of which are
detuned far below the atomic transition frequency $\omega _{a}$, i.e., $%
\left\vert \Delta _{a}\right\vert \equiv \left\vert \omega _{p}-\omega
_{a}\right\vert \gg \Omega _{1,2}$. Adiabatically eliminating the excited
states yields the Hamiltonian of the atom-cavity system%
\begin{equation}
\mathcal{\hat{H}}=-\hbar \left( \Delta _{c}-\frac{g_{c}^{2}}{\Delta _{a}}%
\right) \hat{a}^{\dag }\hat{a}+\int \hat{\psi}^{\dag }(x)\hat{H}_{a}\hat{\psi%
}(x)dx,  \label{HM}
\end{equation}%
with $g_{c}^{2}/\Delta _{a}$ as a constant optical potential per photon and
the cavity detuning $\Delta _{c}=$ $\omega _{p}-\omega _{c}$. The single
particle Hamiltonian density is obtained as (see Appendix A for details)%
\begin{eqnarray}
\hat{H}_{a} &=&-\frac{\hbar ^{2}}{2m}\frac{\partial ^{2}}{\partial x^{2}}%
+\hbar \eta _{1}\cos (kx)(\hat{a}+\hat{a}^{\dag })  \notag \\
&&+\hbar \eta _{2}\sin (kx)(\hat{a}e^{i\theta }+\hat{a}^{\dag }e^{-i\theta })
\notag \\
&&+\hbar V_{1}\cos ^{2}(kx)+\hbar V_{2}\sin ^{2}(kx).  \label{HMA}
\end{eqnarray}%
Here, $\hat{\psi}(x)$ is the matter wave field operator for the atomic
ground state, $\hat{a}$ is the annihilation operator of the cavity photon,
and $k$ is the wave vector of the driving lasers. We have introduced the
driving-field-induced lattice depth $V_{1(2)}=\Omega _{1(2)}^{2}/\Delta _{a}$
and the effective cavity pump strength $\eta _{1(2)}=\Omega
_{1(2)}g_{c}/\Delta _{a}$. The photon loss with rate $\kappa $ is included
in the model via a master equation of the form $\partial _{t}\hat{\rho}%
=-i/\hbar \lbrack \mathcal{\hat{H}},\hat{\rho}]+\mathcal{\hat{L}}\hat{\rho}$%
, where the Lindblad operator acts as $\mathcal{\hat{L}}\hat{\rho}=\kappa (2%
\hat{a}\hat{\rho}\hat{a}^{\dag }-\hat{a}^{\dag }\hat{a}\hat{\rho}-\hat{\rho}%
\hat{a}^{\dag }\hat{a})$. In the following discussion, we neglect the last
two terms of Eq.~(\ref{HMA}) by assuming $V_{1}\approx V_{2}$ for
simplicity. This assumption does not affect the main results of this paper.

As a noteworthy feature of the system, two out-of-phase atomic density
waves, $\cos (kx)$ and $\sin (kx)$, are respectively coupled to\ two
quadratures of the cavity field. The relative coordinate of the two cavity
quadratures is controlled by a coupling angle $\theta $, which quantifies a
rotation of the field distribution in phase space [see Fig.~\ref{setup}(c)
for illustration]. We emphasize that the pump strength and coupling angle
are both competing parameters that determine the interplay between the two
atomic density waves.

In general, the Hamiltonian~(\ref{HMA}) possesses a $%
\mathbb{Z}
_{2}$ symmetry representing its invariance under the transformation $\hat{a}%
\longrightarrow -\hat{a}$ and$\ x\longrightarrow x+\lambda /2$ with $\lambda
=2\pi /k$. Of particular interest is the special case $\theta =\pi /2$,
where the original $%
\mathbb{Z}
_{2}$ symmetry turns into a $%
\mathbb{Z}
_{2}^{(1)}\otimes $\texttt{\ }$%
\mathbb{Z}
_{2}^{(2)}$ double discrete symmetry \cite{PRAP12}, which is composed of two
other transformations%
\begin{eqnarray*}
&&\left( \hat{a}+\hat{a}^{\dag },i\hat{a}-i\hat{a}^{\dag },x\right) \text{ }%
\underrightarrow{\mathcal{T}_{1}}\text{ }\left( -\hat{a}-\hat{a}^{\dag },i%
\hat{a}-i\hat{a}^{\dag },-x+\lambda /2\right) , \\
&&\left( \hat{a}+\hat{a}^{\dag },i\hat{a}-i\hat{a}^{\dag },x\right) \text{ }%
\underrightarrow{\mathcal{T}_{2}}\text{ }\left( \hat{a}+\hat{a}^{\dag },-i%
\hat{a}+i\hat{a}^{\dag },-x\right) .
\end{eqnarray*}%
This symmetry is further enhanced if both $\theta =\pi /2$ and $\eta _{1}=$ $%
\eta _{2}$ are satisfied. In this case, the Hamiltonian is invariant under
the simultaneous spatial transformation $x\longrightarrow x+X$ and the
cavity-phase rotation $\hat{a}\longrightarrow \hat{a}e^{-ikX}$, which yields
a continuous $U(1)$ symmetry associated with the freedom of an arbitrarily
chosen displacement $X$. In the spirit of Landau's theory, it is anticipated
that the aforementioned symmetries should be spontaneously broken by
corresponding phase transitions. However, the dissipative nature plays a
subtle role in the presented system, which prohibits the steady-state phase
transitions associated with the enhanced $%
\mathbb{Z}
_{2}^{(1)}\otimes $\texttt{\ }$%
\mathbb{Z}
_{2}^{(2)}$ and $U(1)$ symmetries. This is because (i) the $%
\mathbb{Z}
_{2}^{(1)}\otimes $\texttt{\ }$%
\mathbb{Z}
_{2}^{(2)}$ symmetry owned by the Hamiltonian is explicitly broken by the
Lindblad operator, and (ii) the dissipation induces extra phase shift for
the cavity photons, preventing the arbitrariness of the value of $X$, which
therefore makes the $U(1)$ symmetry breaking impossible. The physics
demonstrating these points will be detailed in the subsequent sections.

It is worth noting that, moreover, fixing $\theta =\pi /2$ but keeping $\eta
_{1}$ and $\eta _{2}$ as freely controlled parameters is equivalent to its
dual case, namely setting $\eta _{1}=$ $\eta _{2}$ without any constraint on
$\theta $. To see this clearly, let us set $\theta =\pi /2$ and
reparametrize the effective cavity pump strengths by $\eta _{1}=\eta \cos
(\varphi /2)$ and $\eta _{2}=\eta \sin (\varphi /2)$. The single particle
Hamiltonian~(\ref{HMA}) therefore reads%
\begin{eqnarray}
\hat{H}_{a} &=&-\frac{\hbar ^{2}}{2m}\frac{\partial ^{2}}{\partial x^{2}}%
+\hbar \eta \cos (\varphi /2)\cos (kx)(\hat{a}+\hat{a}^{\dag })  \notag \\
&&+\hbar \eta \sin (\varphi /2)(i\hat{a}-i\hat{a}^{\dag })  \notag \\
&&+\hbar V_{1}\cos ^{2}(kx)+\hbar V_{2}\sin ^{2}(kx).  \label{HMA2}
\end{eqnarray}%
Moving into a new gauge by using the transformations $a\longrightarrow
ae^{i\varphi /2}$ and $x\longrightarrow x-\lambda /8$, the Hamiltonian~(\ref%
{HMA2}) exactly reproduces the form of Eq.~(\ref{HMA}),
\begin{eqnarray}
\hat{H}_{a} &=&-\frac{\hbar ^{2}}{2m}\frac{\partial ^{2}}{\partial x^{2}}%
+\hbar \eta \cos (kx)(\hat{a}+\hat{a}^{\dag })  \notag \\
&&+\hbar \eta \sin (kx)(\hat{a}e^{i\varphi }+\hat{a}^{\dag }e^{-i\varphi })
\notag \\
&&+\hbar V_{1}\cos ^{2}(kx)+\hbar V_{2}\sin ^{2}(kx).
\end{eqnarray}%
where $\eta _{1}=$ $\eta _{2}=\eta $ and $\varphi $ plays the role of $%
\theta $. In this sense, if setting $\theta =\pi /2$ (or equivalently $\eta
_{1}=$ $\eta _{2}$), our model shares some similarities with those in Refs.
\cite{PRLML18,SCIN19,DampT1}. However, as will be shown, letting both $%
\theta $ and $\eta _{1,2}$ to be controllable parameters, the proposed model
accommodates more interesting physics which is out of the reach of other
previous proposals.

\section{Mean-field approach}

\label{sec:approach}

In the thermodynamic limit, it is a good approximation to neglect the
quantum correlation between light and matter and thereby treat them as
classical variables. At this mean-field level, the system is described by a
set of coupled equations for the cavity field amplitude $\left\langle \hat{a}%
(t)\right\rangle =$ $\alpha (t)=\left\vert \alpha (t)\right\vert e^{i\phi
(t)}$, and atomic condensate wave function $\left\langle \hat{\psi}%
(x,t)\right\rangle =\sqrt{N}\psi (x,t)=\sqrt{Nn(x,t)}e^{i\tau }$ (see
Appendix B),%
\begin{eqnarray}
i\frac{\partial }{\partial t}\alpha &=&(-\delta _{c}-i\kappa )\alpha +N\eta
_{1}\Theta _{1}+N\eta _{2}e^{-i\theta }\Theta _{2},  \label{EoM1} \\
i\frac{\partial }{\partial t}\psi &=&\left[ -\frac{\hbar }{2m}\frac{\partial
^{2}}{\partial x^{2}}+\eta _{1}\cos (kx)(\alpha +\alpha ^{\ast })\right.
\notag \\
&&\left. +\eta _{2}\sin (kx)(\alpha e^{i\theta }+\alpha ^{\ast }e^{-i\theta
})\right] \psi ,  \label{EoM2}
\end{eqnarray}%
where $N$ is the atom number, $\delta _{c}=\Delta _{c}-g_{c}^{2}/\Delta _{a}$
is the effective cavity detuning, and $\Theta _{1}\equiv \int n(x)\cos
(kx)dx $ and $\Theta _{2}\equiv \int n(x)\sin (kx)dx$ respectively\
represent the occupations of the two out-of-phase density modes, which we
identify as order parameters. The last two terms of Eq.~(\ref{EoM1}) account
for the cavity photon generation rates. Note that these two terms
respectively come from the coherent scattering between the pump field and
different atomic density modes, giving rise to distinct cavity photons. That
is, the term proportional to $\eta _{1}$ excites only one quadrature of the
cavity photons, whereas the other term contributes another quadrature which
is characterized by a rotation of $\theta $ in the phase space. It should be
noticed that these two quadratures of cavity field are basically
nonorthogonal to each other except for $\theta =\pi /2$. The backaction of
the photon scattering on the atomic matter wave is reflected on the terms
proportional to $\cos (kx)$ and $\sin (kx)$ in Eq.~(\ref{EoM2}). These terms
generate a space-dependent optical potential which has a periodicity of $%
\lambda $.

As we are interested in the steady state of the system, we self-consistently
solve Eqs.~(\ref{EoM1})-(\ref{EoM2}) by setting $\partial _{t}\alpha =0$ and
$i\partial _{t}\psi =\mu \psi $, where $\mu $ is the chemical potential of
the condensate. It is clear that, if either one of the pump strengths $\eta
_{1}$ and $\eta _{2}$ is set to be zero, the system reduces to the
conventional transversely pumped BEC inside a cavity, whose physics has been
widely investigated both theoretically \cite{dicketheory3,EPJDD08,PRAP03}
and experimentally \cite{NATK10,SCIR12,PRLJ15}. In that case, by increasing
the pump strength, a \textquotedblleft superradiant phase
transition\textquotedblright\ from a state with no photon inside the cavity
to a state with the appearance of macroscopic cavity field, takes place.
Richer phenomena emerge if both $\eta _{1}$ and $\eta _{2}$ are turned on.
To understand these aspects comprehensively, we first present the result of
closed system ($\kappa =0$) and then inspect the impacts of finite photon
dissipation.

\section{Phase diagram for the closed system}

\label{sec:clophase}

\begin{figure}[tp]
\includegraphics[width=8.0cm]{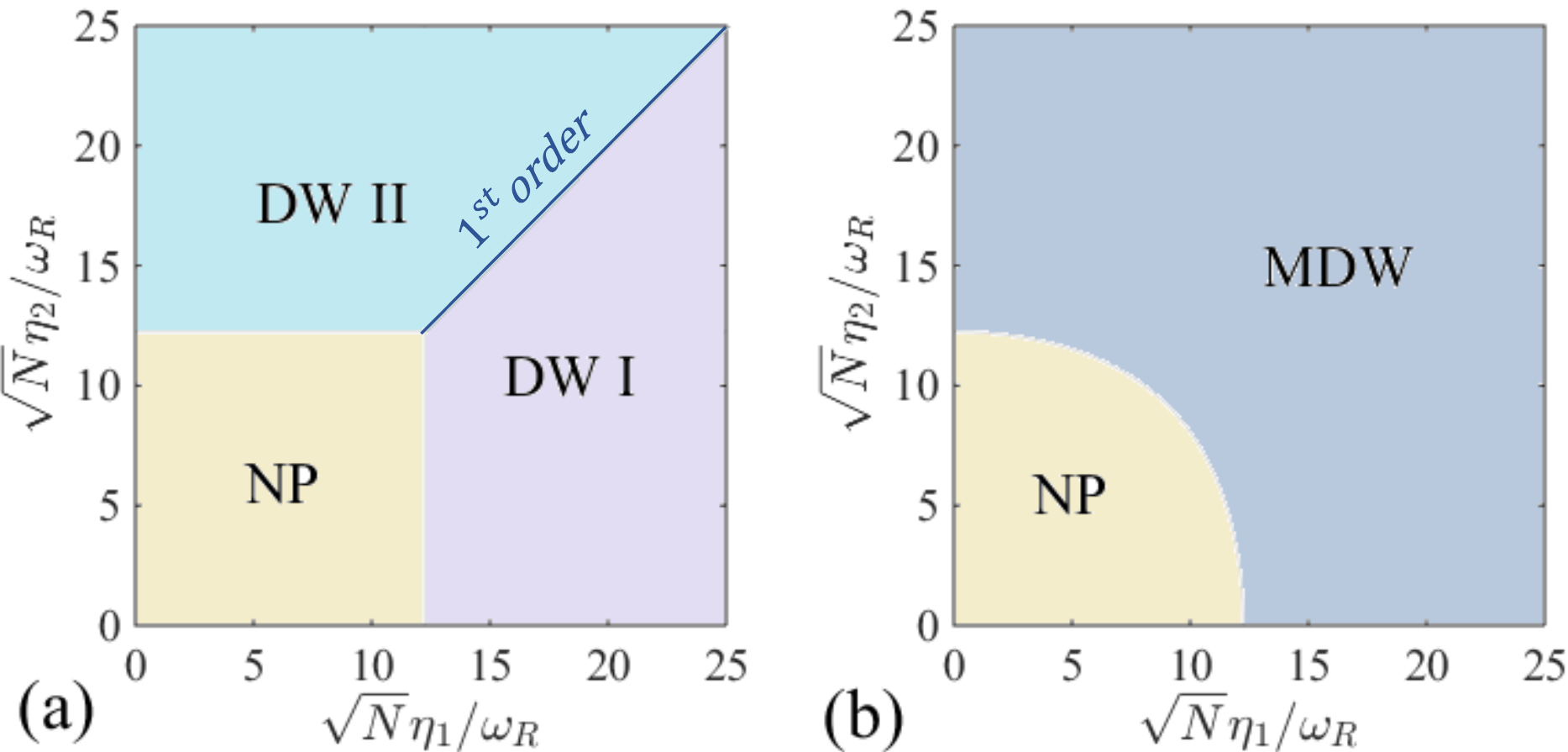}
\caption{Steady-state phase diagrams for closed systems ($\protect\kappa =0$%
) with (a) $\protect\theta =\protect\pi /2$ and (b) $\protect\theta =\protect%
\pi /5$, when $\protect\delta _{c}/\protect\omega _{R}=-300$. }
\label{PhaseI}
\end{figure}

Figure~\ref{PhaseI} plots the phase diagrams for the dissipationless ($%
\kappa =0$) BEC-cavity system as a function of $\eta _{1}$ and $\eta _{2}$.
We first pay attention to the orthogonal coupling case, $\theta =\pi /2$
[see Fig. \ref{PhaseI}(a)], considering its particular symmetry. According
to the values of $\eta _{1}$ and $\eta _{2}$, the steady state is identified
as four different quantum phases. Specifically, when both $\eta _{1}$ and $%
\eta _{2}$ are below a critical value $\eta _{c}=\sqrt{-\delta _{c}\omega
_{R}/2N}$(see Sec.~\ref{sec:stability} for the derivation), the cavity mode
is empty and the density of the condensate keeps uniform with $\Theta
_{1}=\Theta _{2}=0$, corresponding to the normal phase (NP). For $\eta
_{1}>\eta _{c}$ and $\eta _{1}>\eta _{2}$, the BEC is driven into a
self-organized density-wave state characterized by $\Theta _{1}\neq 0$ and $%
\Theta _{2}=0$, which we denote as density wave \textrm{I }(DW \textrm{I}).
Similarly, for $\eta _{2}>\eta _{c}$ and $\eta _{2}>\eta _{1}$, we achieve
another density-wave state characterized by $\Theta _{1}=0$ and $\Theta
_{2}\neq 0$, which is termed density wave\textrm{\ II }(DW \textrm{II}).
Here, the DW \textrm{I }and DW \textrm{II} are essentially symmetry-broken
states which respectively break the $%
\mathbb{Z}
_{2}^{(1)}$ and $%
\mathbb{Z}
_{2}^{(2)}$ symmetries. A more interesting case is $\eta _{1}=\eta _{2}>\eta
_{c}$, where both two density modes are exited with $\Theta _{1}\neq 0$ and $%
\Theta _{2}\neq 0$, and we name this phase as mixed density wave (MDW).
Since in this case, the cavity-field phase $\phi $ can spontaneously take
any arbitrary value between $0$ to $2\pi $, the continuous $U(1)$ symmetry
is broken.

As phase diagrams for any $\theta \neq \pi /2$ resemble each other (they
distinguish themselves solely by minor modifications of the phase
boundaries), we take $\theta =\pi /5$ as a representative example. As shown
in Fig.~\ref{PhaseI}(b), the NP is located within a zone encircled by a
smooth phase boundary. For points $\{\eta _{1}$, $\eta _{2}\}$ outside this
zone, we have $\Theta _{1}\neq 0$ and $\Theta _{2}\neq 0$, corresponding to
the MDW. This picture persists for any coupling angle with $\theta \neq \pi
/2$, implying that a discrepancy from $\theta =\pi /2$ introduces a coupling
between the two density modes $\cos (kx)$ and $\sin (kx)$, and thus excludes
the emergence of both the DW \textrm{I }and DW \textrm{II}. In other words,
the only allowed phase transition is the one from the NP to the MDW.

By further investigating the discontinuities of the order parameters, we
find the transition from the DW \textrm{I }to the DW \textrm{II }is of first
order while the transitions between any other two phases are of second order.

\section{Stability analysis}

\label{sec:stability}

We start to investigate the more appealing driven-dissipative properties by
incorporating a nonzero photon-loss rate $\kappa $ into the model. Since any
potential dissipation-induced instability can not be fully captured by
solely solving the equations of motion, we prefer to carry out a stability
analysis around the trivial solution ($\psi \equiv 1/\sqrt{L}$, $\alpha =0$)
before presenting the final phase diagram. To this end, we work on the
dispersive limit, saying $(\left\vert \delta _{c}\right\vert ,\kappa )\gg
(\omega _{R},\sqrt{N}\eta _{1,2})$ with $\omega _{R}=\hbar k^{2}/2m$ being
the recoil frequency, which allows us to adiabatically eliminate the cavity
field by equating the field amplitude $\alpha $ with its steady-state value $%
\alpha =(N\eta _{1}\Theta _{1}+N\eta _{2}e^{-i\theta }\Theta _{2})/(\delta
_{c}+i\kappa )=R\exp (i\chi )(N\eta _{1}\Theta _{1}+N\eta _{2}e^{-i\theta
}\Theta _{2})$. Note here $R=1/\sqrt{\delta _{c}^{2}+\kappa ^{2}}$ and we
have introduced the dissipation-induced phase shift $\chi =\arctan (\kappa
/\delta _{c})$ \cite{SCIN19}. Under this adiabatic approximation, the
coupled equations of motion reduce to a single one,
\begin{eqnarray}
i\frac{\partial }{\partial t}\psi &\!\!=\!\!\!&\left\{ \!-\frac{\hbar }{2m}%
\frac{\partial ^{2}}{\partial x^{2}}\!+\!\frac{2\hbar N\eta _{1}\cos (kx)}{%
\delta _{c}^{2}+\kappa ^{2}}\left[ \cos (\theta )\delta _{c}\eta
_{2}\left\langle \sin (kx)\right\rangle \right. \right.  \notag \\
&&\left. -\sin (\theta )\kappa \eta _{2}\left\langle \sin (kx)\right\rangle
+\delta _{c}\eta _{1}\left\langle \cos (kx)\right\rangle \right]  \notag \\
&&+\frac{2\hbar N\eta _{2}\sin (kx)}{\delta _{c}^{2}+\kappa ^{2}}\left[ \cos
(\theta )\delta _{c}\eta _{1}\left\langle \cos (kx)\right\rangle \right.
\notag \\
&&\left. \left. -\sin (\theta )\kappa \eta _{1}\left\langle \cos
(kx)\right\rangle +\delta _{c}\eta _{2}\left\langle \sin (kx)\right\rangle
\right] \right\} \psi ,  \label{SEoM}
\end{eqnarray}%
where the symbol $\left\langle \cdot \cdot \cdot \right\rangle $ stands for
the average over single-atom wave function, $\left\langle \psi \right\vert
\cdot \cdot \cdot $ $\left\vert \psi \right\rangle $. We then effect a small
fluctuation from the stationary state ($\psi _{0}$): $\psi (x,t)=e^{-i\mu
t/\hbar }[\psi _{0}(x)+\delta \psi (x,t)]$. Inserting this ansate into Eq.~(%
\ref{SEoM}) and neglecting higher-order correlations, we obtain an equation
linearized in $\delta \psi $,

\begin{widetext}
\begin{eqnarray}
i\frac{\partial }{\partial t}\delta \psi  &=&\left( -\frac{\hbar }{2m}\frac{%
\partial ^{2}}{\partial x^{2}}-\frac{\mu }{\hbar }\right) \delta \psi +\frac{%
2\eta _{1}\cos (kx)}{\delta _{c}^{2}+\kappa ^{2}}\left[ \cos (\theta )\delta
_{c}\eta _{2}(\left\langle \delta \psi \right\vert \sin (kx)\left\vert \psi
_{0}\right\rangle +\left\langle \psi _{0}\right\vert \sin (kx)\left\vert
\delta \psi \right\rangle )\right.   \notag \\
&&\left. -\sin (\theta )\kappa \eta _{2}\left( \left\langle \delta \psi
\right\vert \sin (kx)\left\vert \psi _{0}\right\rangle +\left\langle \psi
_{0}\right\vert \sin (kx)\left\vert \delta \psi \right\rangle \right)
+\delta _{c}\eta _{1}\left( \left\langle \delta \psi \right\vert \cos
(kx)\left\vert \psi _{0}\right\rangle +\left\langle \psi _{0}\right\vert
\cos (kx)\left\vert \delta \psi \right\rangle \right) \right] \psi _{0}
\notag \\
&&+\frac{2\eta _{1}\cos (kx)}{\delta _{c}^{2}+\kappa ^{2}}\left[ \cos
(\theta )\delta _{c}\eta _{2}(\left\langle \delta \psi \right\vert \cos
(kx)\left\vert \psi _{0}\right\rangle +\left\langle \psi _{0}\right\vert
\cos (kx)\left\vert \delta \psi \right\rangle )+\sin (\theta )\kappa \eta
_{2}\times \right.   \notag \\
&&\left. (\left\langle \delta \psi \right\vert \cos (kx)\left\vert \psi
_{0}\right\rangle +\left\langle \psi _{0}\right\vert \cos (kx)\left\vert
\delta \psi \right\rangle )+\delta _{c}\eta _{2}(\left\langle \delta \psi
\right\vert \sin (kx)\left\vert \psi _{0}\right\rangle +\left\langle \psi
_{0}\right\vert \sin (kx)\left\vert \delta \psi \right\rangle )\right] \psi
_{0}.  \label{FEoM}
\end{eqnarray}
\end{widetext}We further assume the fluctuation evolves in the form: $\delta
\psi (x,t)=\delta \psi _{+}(x)e^{-i\omega t/\mathcal{\hbar }}+\delta \psi
_{-}^{\ast }(x)e^{i\omega ^{\ast }t/\mathcal{\hbar }}$, where $\omega =\nu
-i\gamma $ is a complex parameter with $\nu $ and $\gamma $ being the
oscillation frequency and damping rate, respectively. Equation~(\ref{FEoM})
is then recast in a matrix form, $\omega \mathbf{v}=M\mathbf{v}$, where $%
\mathbf{v=}(\delta \psi _{+},\delta \psi _{-})^{\mathbf{T}}$ and%
\begin{equation}
M=\left(
\begin{array}{cc}
H_{0}/\hbar +\Pi _{\ast } & \Pi \\
-\Pi & -H_{0}/\hbar -\Pi _{\ast }%
\end{array}%
\right) ,  \label{FEoM1}
\end{equation}%
with $\Pi =$ $\Xi _{+}\mathcal{I}_{+}+\Xi _{-}\mathcal{I}_{-}$ and $\Pi
_{\ast }=\Xi _{+}\mathcal{I}_{+\ast }+\Xi _{-}\mathcal{I}_{-\ast }$. In the
matrix (\ref{FEoM1}), $H_{0}=$ $-\hbar ^{2}/2m\partial _{x}^{2}-\mu $, $\Xi
_{+}=$ $N[2\eta _{1}^{2}\cos (kx)\delta _{c}\psi _{0}+2\eta _{1}\eta
_{2}\sin (kx)(\cos (\theta )\delta _{c}+\sin (\theta )\kappa )\psi
_{0}]/(\delta _{c}^{2}+\kappa ^{2})$, $\Xi _{-}=$ $N[2\eta _{2}^{2}\sin
(kx)\delta _{c}\psi _{0}+2\eta _{1}\eta _{2}\cos (kx)(\cos (\theta )\delta
_{c}-\sin (\theta )\kappa )\psi _{0}]/(\delta _{c}^{2}+\kappa ^{2})$, and $%
\mathcal{I}_{\pm }$ ($\mathcal{I}_{\pm \ast }$) is an integral operator
defined as $\mathcal{I}_{\pm }\xi =\int_{0}^{\lambda }\psi _{0}(x)\cos
(kx-\pi /4\pm \pi /4)\xi dx/\lambda $ ($\mathcal{I}_{\pm \ast }\xi
=\int_{0}^{\lambda }\psi _{0}^{\ast }(x)\cos (kx-\pi /4\pm \pi /4)\xi
dx/\lambda $). Assuming uniform condensate distribution ($\psi _{0}\equiv 1/%
\sqrt{L}$), the definition of the integral operators $\mathcal{I}_{\pm }$ and%
$\ \mathcal{I}_{\pm \ast }$ indicates that only the Fourier components $\cos
(kx)$ and $\sin (kx)$ couple to the fluctuations, which motivates us to
search solutions in the form
\begin{eqnarray*}
\delta \psi _{+} &\!\!=\!\!&\frac{1}{2}\left[ (\delta \psi _{+}^{1}+\delta
\psi _{-}^{1})\cos (kx)+(\delta \psi _{+}^{2}+\delta \psi _{-}^{2})\sin (kx)%
\right] , \\
\delta \psi _{-} &\!\!=\!\!&\frac{1}{2}\left[ (\delta \psi _{+}^{1}-\delta
\psi _{-}^{1})\cos (kx)+(\delta \psi _{+}^{2}-\delta \psi _{-}^{2})\sin (kx)%
\right] .
\end{eqnarray*}%
Under the basis of $\mathbf{v}^{\prime }=(\delta \psi _{+}^{1}$, $\delta
\psi _{-}^{1}$, $\delta \psi _{+}^{2}$, $\delta \psi _{-}^{2})^{\text{%
\textbf{T}}}$, it is straightforward to write the dynamical matrix as%
\begin{equation}
\mathcal{M}=\left(
\begin{array}{cccc}
0 & \omega _{R} & 0 & 0 \\
\omega _{R}+\zeta _{1} & 0 & \omega _{+} & 0 \\
0 & 0 & 0 & \omega _{R} \\
\omega _{-} & 0 & \omega _{R}+\zeta _{2} & 0%
\end{array}%
\right) ,
\end{equation}%
where $\omega _{+}=2N\eta _{1}\eta _{2}R\cos (\theta +\chi )$, $\omega
_{-}=2N\eta _{1}\eta _{2}R\cos (\theta -\chi )$, $\zeta _{1}=2N\eta
_{1}^{2}R\cos (\chi )$, and $\zeta _{2}=2N\eta _{2}^{2}R\cos (\chi )$. Note
that for later convenience, the entries are intentionally parametrized in
terms of $\chi $ and $R$ instead of the more familiar ones $\kappa $ and $%
\delta _{c}$. Here, $\zeta _{1}$ and $\zeta _{2}$ act as energy shifts,
whereas $\omega _{+}$ and $\omega _{-}$ denote the cavity-mediated couplings
between the two density modes. From the definition of $\omega _{\pm }$, it
is clear that the couplings are generated by the nonorthogonal coupling
angle $\theta $ ($\neq \pi /2$) and the photon dissipation $\chi $ ($\neq 0$%
). That said, the role of dissipation is even more particular since it makes
the two couplings asymmetric ($\omega _{+}\neq \omega _{-}$) and even own
opposite signs ($\omega _{+}\omega _{-}<0$), hinting potential
dissipation-induced instabilities, as will be described below.

\begin{figure}[tp]
\includegraphics[width=8.0cm]{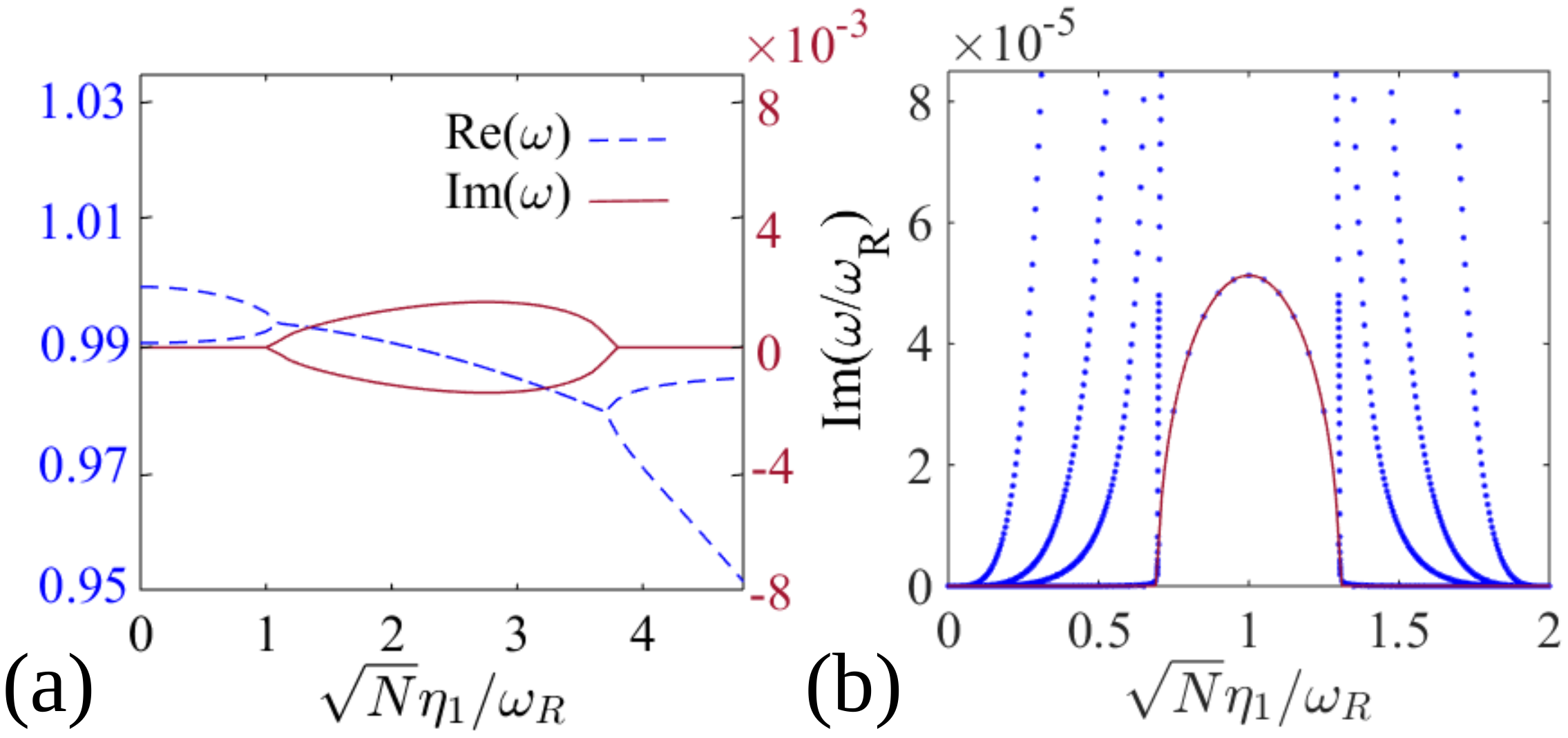} 
\caption{(a) The real and imaginary part of the eigenvalues $\protect\omega $
as a function of $\protect\lambda _{1}$ for $\protect\lambda _{2}/\protect%
\omega _{R}=$ $2$, $\protect\kappa /\protect\omega _{R}=200$, and $\protect%
\delta _{c}/\protect\omega _{R}=-300$. The results are obtained from Eq.~(%
\protect\ref{SPC}). (b) The positive branch of the imaginary part of the
eigenvalues $\protect\omega $ (blue dots), along the transverse cut line, $%
\protect\sqrt{N}\protect\eta _{2}=2\protect\omega _{R}-\protect\sqrt{N}%
\protect\eta _{1}$, depicted by red dotted line in Fig.~\protect\ref{PhaseII}%
(a). The results are obtained by diagonalizing the $6\times 6$ dynamical
matrix including cavity field fluctuations, for $\protect\kappa /\protect%
\omega _{R}=$ $5,15,50,1000,6000$ and $\protect\delta _{c}/\protect\omega %
_{R}=-1.5\protect\kappa $. It can be seen that as the adiabatic limit is
approached, the eigenvalues reduce to the results given by Eq.~(\protect\ref%
{SPC}) (red solid lines).}
\label{spectrum}
\end{figure}

By solving the characteristic equation Det$(\mathcal{M}-\omega I_{4\times
4})=0$, the spectrum of $\mathcal{M}$ is readily obtained as%
\begin{equation}
\omega =\pm \sqrt{\omega _{0}\omega _{R}\pm \frac{\omega _{R}}{2}\sqrt{%
4\omega _{+}\omega _{-}+(\zeta _{1}-\zeta _{2})^{2}}}  \label{SPC}
\end{equation}%
with $\omega _{0}=\omega _{R}+(\zeta _{1}+\zeta _{2})/2$. The zero frequency
($\omega =0$ ) solution of Eq.~(\ref{SPC}) yields the threshold pump
strengths above which the uniform distributed atomic gases self-organize
into density waves. Especially for $\kappa =0$ and $\theta =\pi /2$, the two
pump strengths decouple and we get a simple critical value $\eta _{c}=\sqrt{%
-\delta _{c}\omega _{R}/2N}$. A state becomes dynamically unstable if $%
\omega $ acquires both a positive imaginary part and a nonzero real part. By
inspecting the expression of Eq.~(\ref{SPC}), the relation satisfying this
requirement is found to be $4\omega _{+}\omega _{-}+(\zeta _{1}-\zeta
_{2})^{2}<0$, which, after a substitution of system parameters, results in
the following simple form,%
\begin{equation}
\sin ^{2}(\varphi )>\frac{\cos ^{2}(\chi )}{\sin ^{2}(\theta )},  \label{UT}
\end{equation}%
with $\varphi =2\arctan (\eta _{2}/\eta _{1})$ as we have defined in Sec.~%
\ref{sec:system}. Notice that for this case, the imaginary part of
eigenvalues always come in pairs constituted by negative and positive
branches, which represent damping and amplification, respectively [see Fig.~%
\ref{spectrum}(a)]. It is the appearance of the positive branch, namely the
amplified excitation, that renders the NP unstable. The instability is
characterized by the loss of stationary steady state. In fact, a state which
falls into the unstable regime responds to initial small fluctuations by
undamped limit-cycle oscillations \cite{SCIN19,DampT1,DampT2}.

It can be found from Eq.~(\ref{UT}) that, for a closed system ($\chi =0$),
we have $\cos ^{2}(\chi )/\sin ^{2}(\theta )\equiv 1/\sin ^{2}(\theta
)\geqslant 1$, which invalidates the inequality in Eq.~(\ref{UT}) all the
time. This implies that the dissipation plays the key role in the appearance
of the instability, which is in contrast to some standard cavity-BEC systems
\cite{dicketheory2,dicketheory3,dicketheory4,EPJDD08}. There, the impacts of
dissipation are qualitatively minor since only the phase transition point is
altered without major modification of the phase diagram. Another crucial
knowledge we can infer is that the unstable region in the phase diagram is
feasibly controlled by the coupling angle $\theta $. Actually, tuning $%
\theta $ such that $\sin ^{2}(\theta )<\cos ^{2}(\chi )$, the instability
completely disappears, meaning the whole phase diagram is fully stabilized
irrespective of $\eta _{1}$ and $\eta _{2}$. The equality, $\sin ^{2}(\theta
)=\cos ^{2}(\chi )$, defines a critical point separating a fully stable
regime and a regime with possible instability [see Fig.~\ref{PhaseIV}(a) for
example]. Conversely, the unstable region is maximally enlarged when $\theta
=\pi /2$, which is nothing but the orthogonal coupling case realized in
Refs.~\cite{PRLML18,SCIN19}. From this point of view, embedding a tunable
coupling angle in the light-matter interaction, our proposal offers new
possibilities to either enhance or weaken the dissipation-induced
instability in a controlled manner.

\section{Steady-state quantum phases for the driven-dissipative system}

\label{sec:dissphase}

\begin{figure}[tp]
\includegraphics[width=8.0cm]{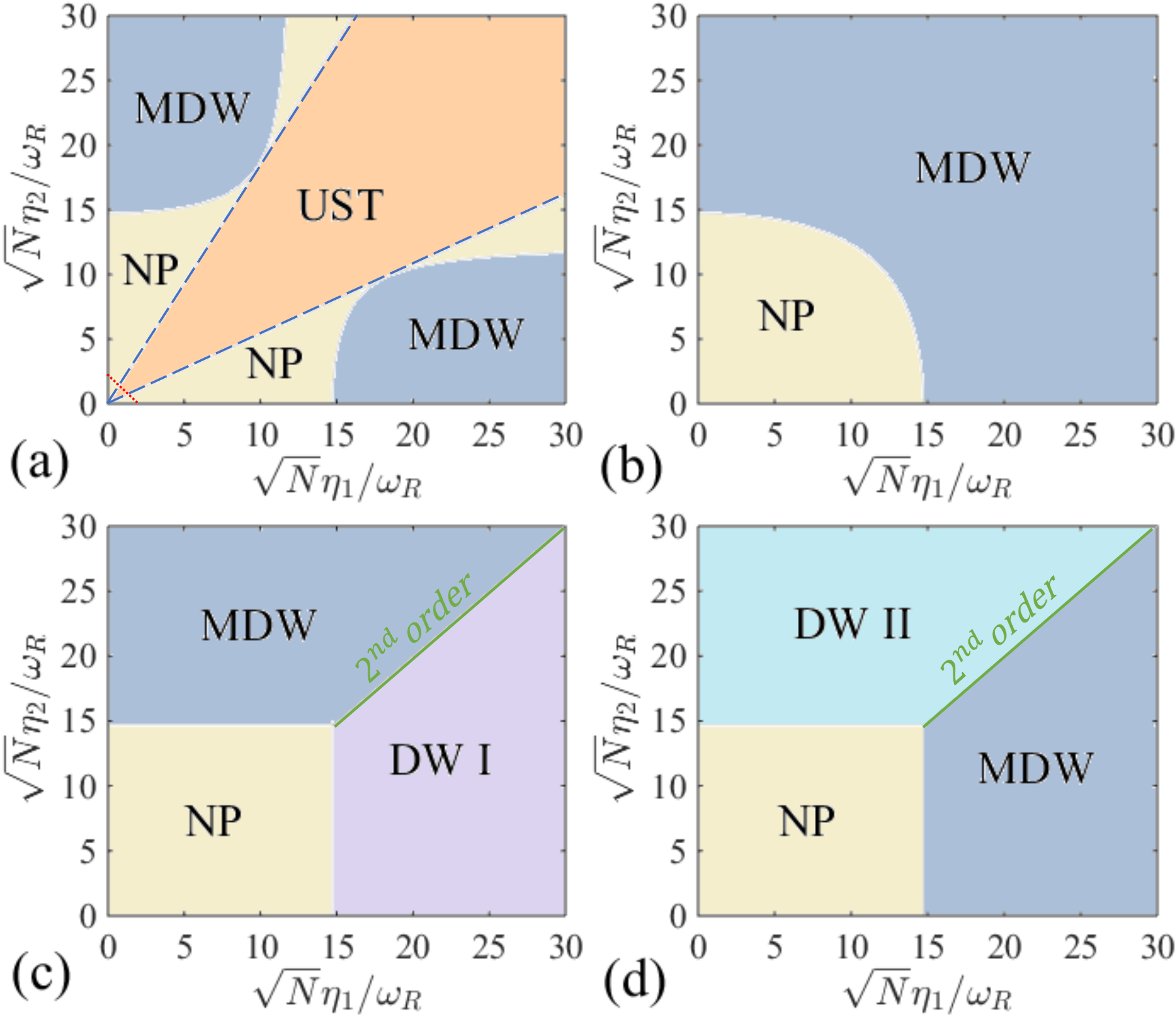}
\caption{Steady-state phase diagrams for the dissipative system, determined
by the equations of motion (\protect\ref{EoM1}-\protect\ref{EoM2}) and a
stability analysis, for varying coupling angles with (a) $\protect\theta =%
\protect\pi /2$, (b) $\protect\theta =\protect\pi /5$, (c) $\protect\theta =-%
\protect\chi +\protect\pi /2\approx 4.12$ and (d) $\protect\theta =\protect%
\chi +\protect\pi /2\approx -0.98$, when $\protect\kappa /\protect\omega %
_{R}=200$ and $\protect\delta _{c}/\protect\omega _{R}=-300$. In (a), the
blue dashed lines are defined by $\sin ^{2}(\protect\varphi )=\cos ^{2}(%
\protect\chi )$.}
\label{PhaseII}
\end{figure}

It is the right stage to explore the quantum phases systematically. Figure~%
\ref{PhaseII} depicts the steady-state phase diagrams for several
representative coupling angles (More phase diagrams and their comparison
with cases of closed system are attributed to Appendix C). We first focus on
the orthogonal coupling case $\theta =\pi /2$. As shown in Fig.~\ref{PhaseII}%
(a), the phase diagram is dramatically distinct from its equilibrium analog
[see Fig.~\ref{PhaseI}(a)]. An immediate observation is that the DW \textrm{I%
} and DW \textrm{II }predicted in Fig.~\ref{PhaseI}(a) are mixed into a MDW
due to the dissipative coupling. Moreover, the expected $U(1)$
symmetry-broken phase transition for $\eta _{1}=\eta _{2}$ vanishes, and a
considerably large region of dynamical instability (UST), enclosed by the
critical curves defined by $\sin ^{2}(\varphi )=\cos ^{2}(\chi )$ (the blue
dashed lines), emerges. As an additional inference, the equal-coupling case
(i.e., $\eta _{1}=\eta _{2}$) is sensitive to the dissipation so much so
that any infinitely small $\kappa $ leads to an instability.

The physics behind this can be well understood in a semi-classical picture.
Treating quantum operators classically, we express the total single-particle
energy as $E=-(\hbar ^{2}/2m)\partial _{x}^{2}+\mathcal{E(}x)$, where the
self-consistent potential is given by
\begin{equation*}
\mathcal{E(}x)=2\left\vert \alpha \right\vert \eta _{1}\cos (\phi )\cos
(kx)-2\left\vert \alpha \right\vert \eta _{2}\sin (\phi )\sin (kx).
\end{equation*}%
The onset of the self-organization is triggered by the periodicity of $%
\mathcal{E(}x)$, attracting more atoms to its minima where the equation $%
\partial _{x}\mathcal{E}=0$ applies. This links the position coordinate with
the cavity phase via%
\begin{equation}
\tan (kx)=-\tan \left( \frac{\varphi }{2}\right) \tan (\phi ).  \label{SX}
\end{equation}%
On the other hand, the steady-state solution of the cavity amplitude reads $%
\alpha =NR$ $e^{i\chi }[\eta _{1}\cos (kx)-i\eta _{2}\sin (kx)]$, producing%
\begin{equation}
\tan (\phi )=\frac{\sin (\chi )-\cos (\chi )\tan (\varphi /2)\tan (kx)}{\cos
(\chi )+\sin (\chi )\tan (\varphi /2)\tan (kx)}.  \label{SPhi}
\end{equation}%
The existence of a solution for Eqs.~(\ref{SX}) and (\ref{SPhi}) requires $%
\sin (\varphi )>\cos (\chi )$, which agrees with the result obtained from
the stability analysis. This picture also explains the absence of the $U(1)$
symmetry breaking for the case $\eta _{1}=\eta _{2}$ (i.e., $\tan (\varphi
/2)=1$), since the dissipation-induced phase shift $\chi $ imposes extra
constraint on the degree of freedom of $\phi $ through Eq.~(\ref{SPhi}),
which makes it frozen to specific value instead of picking up a random
number from $0$ to $2\pi $.

\begin{figure}[tp]
\includegraphics[width=8.0cm]{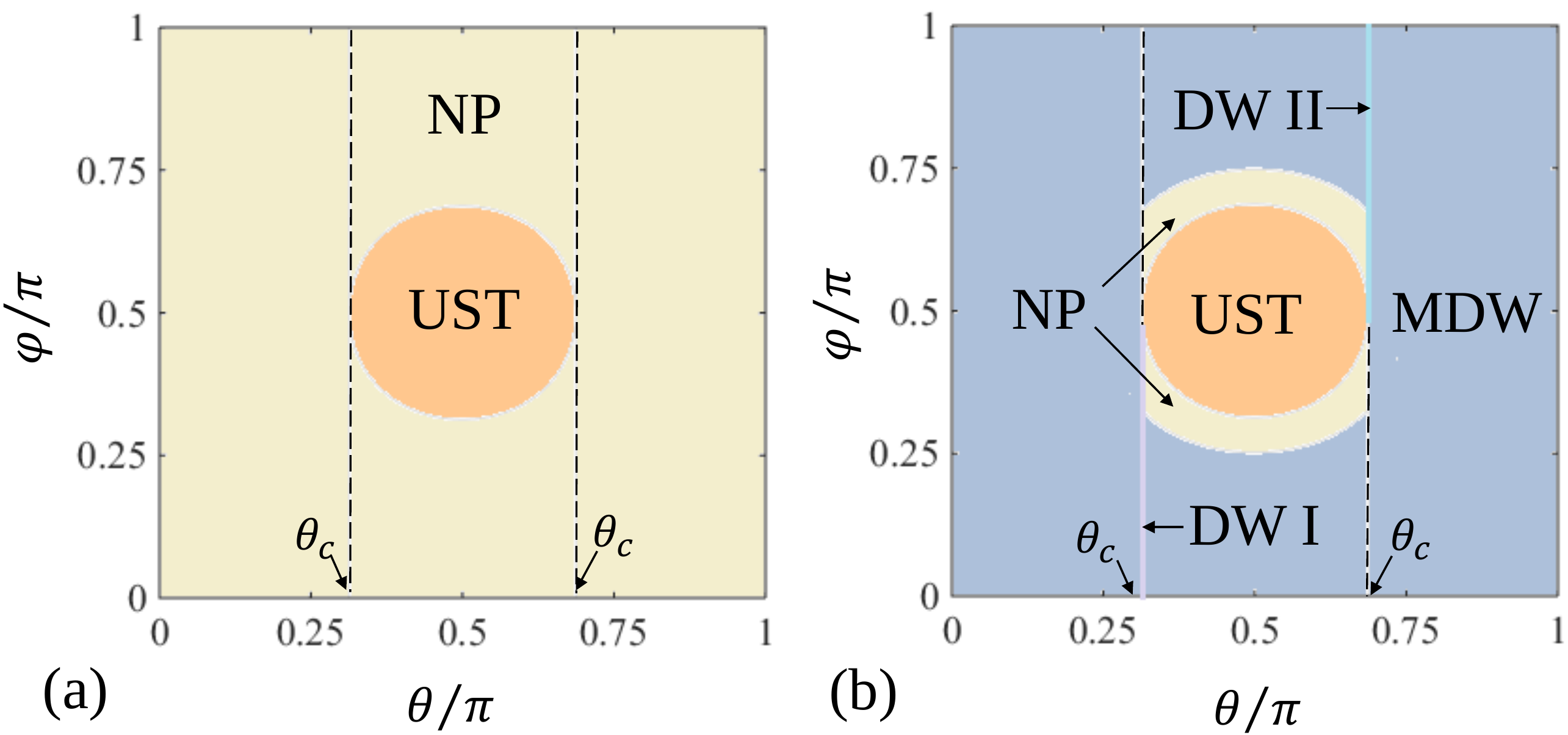}
\caption{Steady-state phase diagrams in the rescaled parameter space $\{%
\protect\theta /\protect\pi ,\protect\varphi /\protect\pi \}$ for (a) $%
\protect\sqrt{N}\protect\eta /\protect\omega _{R}=10$ and (b) $\protect\sqrt{%
N}\protect\eta /\protect\omega _{R}=30$, when $\protect\kappa /\protect%
\omega _{R}=200$ and $\protect\delta _{c}/\protect\omega _{R}=-300$. Region
UST represents dynamically unstable phase, and the black dashed lines are
defined by $\sin ^{2}(\protect\theta )=\cos ^{2}(\protect\chi )$, which
determines the critical coupling angle $\protect\theta _{c}$.}
\label{PhaseIV}
\end{figure}

Along this reasoning, it is expected that phase diagrams for other coupling
angles should be qualitatively similar, saying the self-organized phase
cannot be anything but the MDW [see Fig.~\ref{PhaseII}(b) for example].
However, an intriguing phenomenon occurs when situating $\theta $ at the
critical points described by $\sin ^{2}(\theta )=\cos ^{2}(\chi )$ (i.e., $%
\theta =\theta _{c}=\pm \chi \pm \pi /2$), as shown in Figs.~\ref{PhaseII}%
(c) and \ref{PhaseII}(d). Considering the duality of Figs.~\ref{PhaseII}(c)
and \ref{PhaseII}(d), let us take $\theta =-\chi \pm \pi /2$ as an example.
In this case, the phase diagram exactly recovers the skeleton of that in
Fig.~\ref{PhaseI}(a) where a closed system with $\theta =$ $\pi /2$
operates. That is to say, the whole phase diagram is divided into three
different regions, $\{\eta _{1}\leqslant \tilde{\eta}_{c}$, $\eta
_{2}\leqslant \tilde{\eta}_{c}\}$, $\{\eta _{1}>\tilde{\eta}_{c}$, $\eta
_{1}>\eta _{2}\}$, and $\{\eta _{2}>\tilde{\eta}_{c}$, $\eta _{2}>\eta
_{1}\} $ with a redefined critical pump strength $\tilde{\eta}_{c}=\eta
_{c}/\sin (\theta )$. Nevertheless, the major difference lies in the region $%
(\eta _{2}>\tilde{\eta}_{c}$, $\eta _{2}>\eta _{1})$ where the MDW
supersedes the DW \textrm{II}, and the first order transition presented in
Fig.~\ref{PhaseI}(a) becomes second order here. As complements, Figs.~\ref%
{PhaseIV}(a) and \ref{PhaseIV}(b) show phase diagrams in the $\theta
-\varphi $ plane for different pump strengths $\eta $ $(\equiv \sqrt{\eta
_{1}^{2}+\eta _{2}^{2}}) $, from which the particularity of $\theta _{c}$
becomes clearer. These results look a bit counterintuitive, since both the
nonorthogonal coupling and the cavity dissipation are apt to mix the two
density modes. Our finding shows that the dissipation defines a particular
coupling angle $\theta _{c}=\pm \chi \pm \pi /2$, in which the two mixing
elements cooperate and somehow counteract each other.

Let us give a description for this exotic behavior. Observing only the
Fourier components $\cos (kx)$ and $\sin (kx)$ of a fluctuation of the
condensate wave function can excite a nonzero cavity field, we construct a
trial initial wave function $\psi (x,0)=\sqrt{1/L}+\epsilon _{1}\sqrt{2/L}%
\cos (kx)+\epsilon _{2}\sqrt{2/L}\sin (kx)$, with $\left\vert \epsilon
_{1,2}\right\vert \ll 1$ \cite{EPJDD08}. Propagating $\psi (x,0)$ by one
iteration step of the imaginary time $\Delta \tau $ ($\tau =it$), we have $%
\psi (x,\Delta \tau )=$ $\sqrt{1/L}+\delta \psi (x,\Delta \tau )$, where%
\begin{eqnarray}
\delta \psi (x,\Delta \tau ) &&\!=\left\{ \epsilon _{1}-\left[ 2NR\cos (\chi
-\theta )\eta _{1}\eta _{2}\epsilon _{2}+\omega _{R}\epsilon _{1}\right.
\right. \!\!  \notag \\
&&\left. \left. +2NR\cos (\chi )\eta _{1}^{2}\epsilon _{1}\right] \Delta
\tau \right\} \sqrt{\frac{2}{L}}\cos (kx)  \notag \\
&&+\left\{ \epsilon _{2}-\left[ 2NR\cos (\chi +\theta )\eta _{1}\eta
_{2}\epsilon _{1}+\omega _{R}\epsilon _{2}\right. \right.  \notag \\
&&\left. \left. +2NR\cos (\chi )\eta _{2}^{2}\epsilon _{2}\right] \Delta
\tau \right\} \sqrt{\frac{2}{L}}\sin (kx).  \label{DPsi1}
\end{eqnarray}%
Under the basis of $\mathbf{v}^{\prime \prime }=(\sqrt{2/L}\cos (kx)$, $%
\sqrt{2/L}\sin (kx))$, Eq.~(\ref{DPsi1}) can be formulated in the matrix
form, $\boldsymbol{\delta \psi }(x,\Delta \tau )=(\delta \psi _{1}(x,\Delta
\tau ),\delta \psi _{2}(x,\Delta \tau ))^{\text{\textbf{T}}}=\Gamma
(\epsilon _{1},\epsilon _{2})^{\text{\textbf{T}}}$, where
\begin{equation}
\Gamma =\left(
\begin{array}{cc}
1-\mathcal{D}_{1}\Delta \tau & \mathcal{N}_{-}\Delta \tau \\
\mathcal{N}_{+}\Delta \tau & 1-\mathcal{D}_{2}\Delta \tau%
\end{array}%
\right) ,
\end{equation}%
with $\mathcal{D}_{1,2}=2NR\cos (\chi )\eta _{1,2}^{2}+\omega _{R}$ and $%
\mathcal{N}_{\pm }=-2NR\cos (\chi \pm \theta )\eta _{1}\eta _{2}$. Inserting
$\theta =-\chi +\pi /2$ into $\Gamma $ and diagonalizing it, we get two
eigenvalues $\tilde{\Omega}_{1}=1-[2NR\cos (\chi )\eta _{1}^{2}+\omega
_{R}]\Delta \tau $ and $\tilde{\Omega}_{2}=1-[2NR\cos (\chi )\eta
_{2}^{2}+\omega _{R}]\Delta \tau $, whose eigenvectors respectively reads $%
\mathbf{v}_{1}=(1,0)^{\text{T}}$ and $\mathbf{v}_{2}=(-2\eta _{1}\eta
_{2}\sin (\chi )/(\eta _{1}^{2}-\eta _{2}^{2}),1)^{\text{T}}$. Utilizing $%
\tilde{\Omega}_{1,2}$ and $\mathbf{v}_{1,2}$, it is straightforward to
obtain the wave function at $n\Delta \tau $,%
\begin{eqnarray}
\psi (x,n\Delta \tau ) &=&\sqrt{\frac{1}{L}}+\epsilon _{2}\sqrt{\frac{2}{L}}%
\left[ \sin (kx)\right.  \notag \\
&&\left. -\frac{2\eta _{1}\eta _{2}\sin (\chi )}{\eta _{1}^{2}-\eta _{2}^{2}}%
\cos (kx)\right] \tilde{\Omega}_{2}^{n}  \notag \\
&&+\epsilon _{1}^{\prime }\sqrt{\frac{2}{L}}\cos (kx)\tilde{\Omega}_{1}^{n},
\label{PsiN}
\end{eqnarray}%
where $\epsilon _{1}^{\prime }=$ $\epsilon _{1}+2\eta _{1}\eta _{2}\sin
(\chi )/(\eta _{1}^{2}-\eta _{2}^{2})\epsilon _{2}$ and $n$ can be any
integer number. In Eq.~(\ref{PsiN}), $\tilde{\Omega}_{1,2}<1$ ($\tilde{\Omega%
}_{1,2}>1$) represents decay (amplification) of corresponding modes, leading
to the normal (self-organized) state in the long-time limit. Notice that the
second line of Eq.~(\ref{PsiN}) involves a term proportional to $\sin
(kx)-2\eta _{1}\eta _{2}\sin (\chi )/(\eta _{1}^{2}-\eta _{2}^{2})\cos (kx)$%
, it thus becomes evident that for $\eta _{1}>\tilde{\eta}_{c}$ and $\eta
_{2}<\tilde{\eta}_{c}$ (namely, $\tilde{\Omega}_{1}>1$ and $\tilde{\Omega}%
_{2}<1$), only the cosinelike density wave $\propto \cos (kx)$\ emerges (DW
\textrm{I}), while for $\eta _{2}>\tilde{\eta}_{c}$ and $\eta _{1}<\tilde{%
\eta}_{c}$ (namely, $\tilde{\Omega}_{1}<1$ and $\tilde{\Omega}_{2}>1$), both
two density waves are simultaneously excited (MDW). We emphasize that the
above derivation is mainly based on a perturbation assumption, which works
only around weak excitation regime, it should therefore not be strange that
the present framework is not able to precisely predict the phase boundary
between DW \textrm{I }and MDW.

For completeness, we put diagrams of the order parameters $\Theta _{1}$ and $%
\Theta _{2}$, from which one obtain the phase diagrams of Figs.~\ref{PhaseI}
and \ref{PhaseII}, in Appendix D.

\section{Beyond adiabatic elimination}

\label{sec:nonadiabaticity}

Up to now, the discussion is restricted to the adiabatic limit where
fluctuations of the cavity amplitude is omitted. We now go beyond the
adiabatic approximation by including the dynamics of the cavity fluctuations
$\delta \alpha $ and $\delta \alpha ^{\ast }$ (see Appendix E). By doing
this, we get a $6\times 6$ dynamical\ matrix whose spectrum can not be
expressed analytically. The numerical diagonalization of this matrix
suggests that, the nonadiabaticity exerts no influence on the self-organized
phase but makes the NP unstable for all $\theta \neq 0,\pm \pi $. This
arises from the observation that a nonzero positive imaginary part of the
eigenvalues appears throughout the NP except for $\theta =0,\pm \pi $.
Figure~\ref{spectrum}(b) depicts the imaginary part of the these eigenvalues
for some different $\delta _{c}$ and $\kappa $. We find that approaching the
adiabatic limit $(\left\vert \delta _{c}\right\vert ,\kappa )\gg (\omega
_{R},\sqrt{N}\eta _{1,2})$, the results reduce to that given by Eq.~(\ref%
{SPC}).

\section{Three-mode approximation for the BEC}

\label{sec:threemode}

\begin{figure}[tp]
\includegraphics[width=8.0cm]{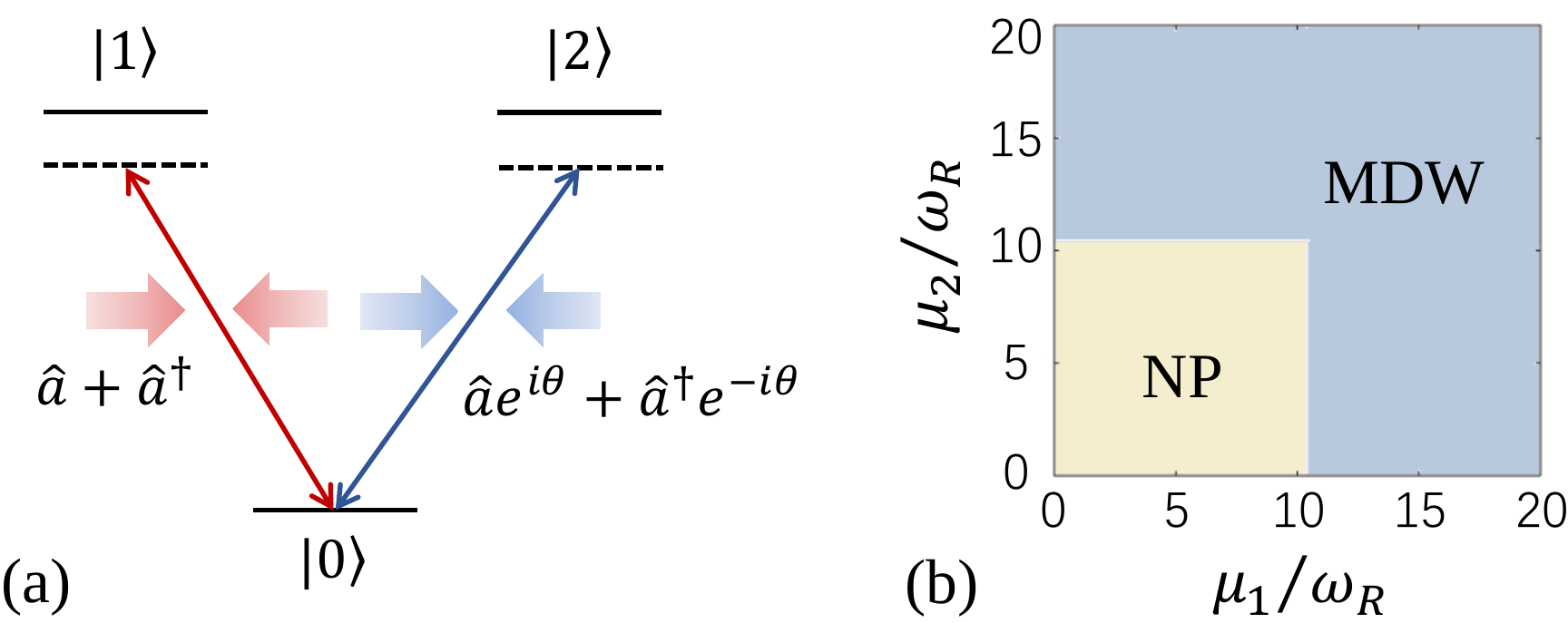}
\caption{(a) Three-level atoms interact with different quadratures of a
single-mode quantized light field via transition channels $\left\vert
0\right\rangle \longleftrightarrow \left\vert 1\right\rangle $ and $%
\left\vert 0\right\rangle \longleftrightarrow \left\vert 2\right\rangle $,
respectively. (b) Phase diagram for the effective model under three-mode
approximation. The parameters are the same as those in Fig.~\protect\ref%
{PhaseII}(c).}
\label{PhaseIII}
\end{figure}

Following the commonly used two-mode approximation \cite%
{NATJ17,NATK10,SCIR12}, the matter field in our model can be spanned by
minimally three Fourier-modes within the single recoil scattering limit,
\begin{equation}
\hat{\psi}(x)=\sqrt{\frac{1}{L}}\left[ \hat{c}_{0}+\hat{c}_{1}\sqrt{2}\cos
(kx)+\hat{c}_{2}\sqrt{2}\sin (kx)\right] ,
\end{equation}%
where $\hat{c}_{0}$, $\hat{c}_{1}$, and $\hat{c}_{2}$ are bosonic
annihilation operators for corresponding modes. It is more convenient to
introduce the collective three-level operator $\hat{\Xi}_{ij}=\sum_{k=1}^{N}%
\left\vert i\right\rangle _{k}\left\langle j\right\vert _{k}$ with atomic
states $\left\{ \left\vert 0\right\rangle _{k},\left\vert 1\right\rangle
_{k},\left\vert 2\right\rangle _{k}\right\} $ $(k=1,2,...,N)$. The operators
$\hat{\Xi}_{ij}$ fulfill the $U(3)$ algebra commutation relations $[\hat{\Xi}%
_{ij}$, $\hat{\Xi}_{kl}]=\delta _{jk}\hat{\Xi}_{il}-\delta _{il}\hat{\Xi}%
_{kj}$. By invoking a generalized-Schwinger representation \cite{RMPK91}, $%
\hat{\Xi}_{ij}=\hat{c}_{i}^{\dagger }\hat{c}_{j}$ $(i,j=0,1,2)$, the
Hamiltonian~(\ref{HM}) in the three-mode subspace reads%
\begin{eqnarray}
\hat{H} &=&-\hbar \delta _{c}\hat{a}^{\dag }\hat{a}-\hbar \omega _{R}\hat{\Xi%
}_{00}+\frac{\hbar \mu _{1}}{\sqrt{N}}(\hat{\Xi}_{01}+\hat{\Xi}_{10})(\hat{a}%
+\hat{a}^{\dag })  \notag \\
&&+\frac{\hbar \mu _{2}}{\sqrt{N}}(\hat{\Xi}_{02}+\hat{\Xi}_{20})(\hat{a}%
e^{i\theta }+\hat{a}^{\dag }e^{-i\theta }),  \label{THAM}
\end{eqnarray}%
with the collective coupling strength $\mu _{1,2}=\eta _{1,2}\sqrt{2N}/2$.
It is easy to check that the symmetry property here follows that in the
Hamiltonian~(\ref{HM}). Especially, when $\mu _{1}=\mu _{2}$ and $\theta
=\pi /2$, the emergent $U(1)$ symmetry is characterized by a conserved
quantity $\mathcal{\hat{C}}=\hat{a}^{\dag }\hat{a}+i(\hat{\Xi}_{12}-\hat{\Xi}%
_{21})$,\ satisfying $[\mathcal{\hat{C}},\hat{H}]=0$. The effective
Hamiltonian~(\ref{THAM}) describes a single-mode quantized light field
interacting with three-level atoms, whose transition channels, $\left\vert
0\right\rangle \longleftrightarrow \left\vert 1\right\rangle $ and $%
\left\vert 0\right\rangle \longleftrightarrow \left\vert 2\right\rangle $,
are coupled by different quadratures of light [see Fig.~\ref{PhaseIII}(a)].

The quantum phases for this model are classified by the expectation values
of $\hat{\Xi}_{11}$ and $\hat{\Xi}_{22}$, whose roles are the same as those
of $\Theta _{1}$ and $\Theta _{2}$, respectively. Similarly, the phase
diagram is straightforwardly obtained by\ exploiting the steady state of the
equations of motion, $i\hbar \partial _{t}\left\langle \hat{\Xi}%
_{ij}\right\rangle =\left\langle [\hat{\Xi}_{ij},\mathcal{\hat{H}}%
]\right\rangle $ and $i\hbar \partial _{t}\left\langle \hat{a}\right\rangle
=\left\langle [\hat{a},\mathcal{\hat{H}}]\right\rangle -i\hbar \kappa
\left\langle \hat{a}\right\rangle $ (see Appendix F for details). While for
most parameters we are interested in, the solutions are in accordance with
the results obtained by directly solving Eqs.~(\ref{EoM1})-(\ref{EoM2}), a
remarkable exception appears when tuning the coupling angle to the critical
values $\theta _{c}=\pm \chi \pm \pi /2$. In this case, the three-level
model predicts only two possible phases: NP and MDW, as shown in Fig.~\ref%
{PhaseIII}(b). This sharply contrasts with Figs.~\ref{PhaseII}(c) and \ref%
{PhaseII}(d), which are plotted based on the solutions for Eqs.~(\ref{EoM1}%
)-(\ref{EoM2}). As a matter of fact, provided the photon dissipation is
incorporated, the three-level model always excludes the emergence of the DW
\textrm{I }and DW \textrm{II}. This finding provides an interesting example
where the effectiveness of the three-mode approximation is radically broken
by the dissipative nature. It is thus a hint that the effective model under
low-excitation-mode approximation may be insufficient in capturing certain
physics when the dissipation starts to play a role. We leave the exploration
of its microscopic origin to the future work.

\section{Experimental consideration}

\label{sec:experiment}

In the proposed experiment, the two driving lasers can be respectively
chosen as left- and right-circularly polarized. Accordingly, the atomic
internal ground and excited states are hyperfine Zeeman states with magnetic
levels $m=0$ and $m\pm 1$, respectively. Given this, a promising candidate
for the phase retarder is the Faraday rotator \cite{JJAP80}, which\ can
impart arbitrary phase difference between the two backreflected
circularly-polarized lasers. Since the coupling angle $\theta $ is acquired
just from the phase retarder, it can be feasibly controlled by simply
varying the magnetic field in the Faraday rotator. Moreover, the realization
of the cosinelike and sinelike density coupling in the Hamiltonian~(\ref{HMA}%
) can be easily achieved by locking the phase difference of the two incident
lasers to be $\pi /2$ (see Appendix A). While the experiment technique to
directly distinguish the two density patterns $\cos (kx)$ and $\sin (kx)$
has been developed \cite{PRLY19,PRAY19}, a more convenient way is to exploit
the one-to-one correspondence between the cavity phase $\phi $ and the
atomic density wave order parameters $\Theta _{1,2}$. In recognition of
this, the goal to identify different density waves is mapped into detecting
the cavity phase, which can be readily accomplished by using a heterodyne
detection system analyzing the light field leaking from the cavity \cite%
{NCR15,PRLR18,SCIN19,PRAW93}.

We then provide a brief estimation of the system parameters based on the
current experimental conditions with $^{87}$Rb atoms \cite%
{NATR16,PRLR18,PRLML18,DampT1}. For laser wavelength $\lambda $ near $780$
nm, the recoil frequency $\omega _{R}$ is estimated to be $\sim 10$ kHz. The
number of trapped atoms which is on the order of $N$ $\sim 10^{4}$ appears
to be practical \cite{NATR16,PRLML18}.\ The atomic detuning can be chosen as
$\Delta _{a}$ $\sim 100$ GHz \cite{PRLR18}, and the parameters $(\left\vert
\Omega _{1,2}\right\vert ,\left\vert g\right\vert ,\left\vert \Delta
_{c}\right\vert ,\kappa )$ are on the order of a few MHz. Thus, the
condition for the adiabatic elimination of the excited atomic levels, saying
$\left\vert \Delta _{a}\right\vert $ $\gg (\left\vert \Omega
_{1,2}\right\vert ,\left\vert g\right\vert ,\left\vert \Delta
_{c}\right\vert )$, is well satisfied. Under this parameters setting, the
collective coupling strengths $\sqrt{N}\eta _{1}$ and $\sqrt{N}\eta _{2}$
can be widely tuned ranging from $0$ to the order of MHz, implying the
self-orgnization condition $\eta _{1}(\eta _{1})\geqslant $ $\eta _{c}$ is
achievable. Furthermore, by properly seting the Rabi frequencies and cavity
detuning, it is easy to place the system in the adiabatic limit of the
cavity field [$(\left\vert \delta _{c}\right\vert ,\kappa )\gg (\omega _{R},%
\sqrt{N}\eta _{1,2})$].

\section{Conclusions}

\label{sec:conclusion}

In summary, we have proposed an experimental scheme, where two density-wave
degrees of freedom of the BEC are coupled to two quadratures of the cavity
field. Different from previous studies, here the coupling angle between the
two quadratures is experimentally tunable, leading to new physics emerging
from nonorthogonal quadratures coupling between light and matter. For a
closed system without dissipation, the two atomic density modes can be
excited respectively by varying the pump strength and coupling angle. This
gives rise to four possible quantum phases, all of which are shown to be
stable against fuctuations. The cavity dissipation, however, plays a
significant role in determining the steady-state phase diagram. For one
thing, it induces a novel unstable region above the normal phase. For the
other, it defines a particular coupling angle, across which the system
exhibits some properties resembling its equilibrium analog. While additional
antidampings may be generated by the nonadiabaticity of the cavity field,
which renders the normal phase unstable, it turns out to be negligibly small
for typical parameters in the current experiments. Moreover, for some
special parameters, the commonly used low-excitation-mode approximation is
shown to be questionable for our model due to the dissipative nature of the
system.

\acknowledgments

This work is supported partly by the National Key R\&D Program of China
under Grant No.~2017YFA0304203; the NSFC under Grants No.~11674200 and
No.~11804204; and 1331KSC.

\vbox{\vskip1cm} \appendix

\section{Effective Hamiltonian}

In this section, we provide the detailed derivation of Hamiltonian~(\ref{HM}%
) in the main text. We start by considering the coupling of internal states
of a single atom, as illustrated in Fig.~\ref{setup}(b) in the main text.
The Hamiltonian can be decomposed as $\hat{H}=\hat{H}_{\text{0}}+%
\overleftarrow{\hat{H}_{\text{I}}}+\overrightarrow{\hat{H}_{\text{I}}}$,
where%
\begin{equation}
\hat{H}_{\text{0}}=\omega _{c}\hat{a}^{\dag }\hat{a}+\sum_{j=1,2}\omega
_{j}\left\vert j\right\rangle \left\langle j\right\vert +\frac{\mathbf{\hat{p%
}}^{2}}{2m}+V_{R}(\mathbf{r}),  \label{SH0}
\end{equation}%
\begin{equation}
\overleftarrow{\hat{H}_{\text{I}}}=-\frac{1}{2}\sum_{j=1,2}\left(
\overleftarrow{\Omega _{j}}(x)e^{-i\omega _{p}t}\left\vert 0\right\rangle
\left\langle j\right\vert +g_{c}\hat{a}\left\vert 0\right\rangle
\left\langle j\right\vert +\text{H.c.}\right) ,  \label{SHL}
\end{equation}%
\begin{equation}
\overrightarrow{\hat{H}_{\text{I}}}=-\frac{1}{2}\sum_{j=1,2}\left(
\overrightarrow{\Omega _{j}}(x)e^{-i\omega _{p}t}\left\vert 0\right\rangle
\left\langle j\right\vert +g_{c}\hat{a}\left\vert 0\right\rangle
\left\langle j\right\vert +\text{H.c.}\right) ,  \label{SHR}
\end{equation}%
with the Rabi frequencies $\overleftarrow{\Omega _{j}}(x)=\Omega _{j}\exp
[i(kx+\vartheta _{j}+\theta _{j})]$ and $\overrightarrow{\Omega _{j}}%
(x)=\Omega _{j}\exp [-i(kx+\vartheta _{j}-\theta _{j})]$. Note that $\hat{H}%
_{\text{0}}$ is the free Hamiltonian and $\overleftarrow{\hat{H}_{\text{I}}}$
($\overrightarrow{\hat{H}_{\text{I}}}$)\ represents the light-matter
interaction contributed by the incident (backreflected) pumping lasers. In
the Hamiltonians~(\ref{SH0})-(\ref{SHR}), $\mathbf{\hat{p}}^{2}/2m$ and $%
V_{R}(\mathbf{r})$ are the kinetic energy and transverse trapping potential
respectively, and $\omega _{j}$ denotes the eigenfrequency of the atomic
state $\left\vert j\right\rangle $ ($j=1,2$). The field operator $\hat{a}$
describes the annihilation of a cavity photon with the frequency $\omega
_{c} $. The transitions $\left\vert 0\right\rangle \leftrightarrow
\left\vert 1\right\rangle $ and $\left\vert 0\right\rangle \leftrightarrow
\left\vert 2\right\rangle $ are respectively driven by two
orthogonally-polarized pumping lasers with the Rabi amplitudes $\Omega _{1}$
and $\Omega _{2}$. H.c. denotes the Hermitian conjugation. Since the BEC is
arranged to be orthogonal to the cavity axis, the atom-cavity coupling $%
g_{c} $ is space independent. We emphasize that the phase of the incident
(backreflecting) laser mediating the transition $\left\vert 0\right\rangle
\leftrightarrow \left\vert j\right\rangle $ is given by $\vartheta
_{j}+\theta _{j}$ ($\vartheta _{j}-\theta _{j}$). Therefore, the phase shift
imparted by the phase retarder for the corresponding transition is $2\theta
_{j}$.

We introduce a time-dependent unitary transformation, $\hat{U}(t)=\exp
[i(\sum_{j=1,2}\left\vert j\right\rangle \left\langle j\right\vert +\hat{a}%
^{\dag }\hat{a})\hbar \omega _{p}t]$, under which the Hamiltonian $\hat{H}$
becomes
\begin{eqnarray}
\hat{H} &=&-\Delta _{c}\hat{a}^{\dag }\hat{a}+\frac{\mathbf{\hat{p}}^{2}}{2m}%
+V_{R}(\mathbf{r})  \notag \\
&&-\sum_{j=1,2}\left( \Delta \left\vert j\right\rangle \left\langle
j\right\vert +\frac{\overleftarrow{\Omega _{j}}(x)}{2}\left\vert
0\right\rangle \left\langle j\right\vert \right.  \notag \\
&&\left. +\frac{\overrightarrow{\Omega _{j}}(x)}{2}\left\vert 0\right\rangle
\left\langle j\right\vert +g\hat{a}\left\vert 0\right\rangle \left\langle
j\right\vert +\text{H.c.}\right) ,  \label{SH1}
\end{eqnarray}%
where $\Delta _{c}=$ $\omega _{p}-\omega _{c}$ is the cavity detuning, $%
\Delta _{a}=$ $\omega _{p}-\omega _{1}\approx \omega _{p}-\omega _{2}$
denotes the detuning between pumping lasers and atomic eigenfrequencies. We
work in the limit of large detuning $\left\vert \Delta _{a}\right\vert $ $%
\gg (\left\vert \Omega _{1,2}\right\vert ,\left\vert g\right\vert
,\left\vert \Delta _{c}\right\vert )$, which allows us to adiabatically
eliminate the excited states $\left\vert 1\right\rangle $ and $\left\vert
2\right\rangle $. The resulting effective Hamiltonian is given as
\begin{eqnarray}
\hat{H} &=&-\delta _{c}\hat{a}^{\dag }\hat{a}+\frac{\mathbf{\hat{p}}^{2}}{2m}%
+V_{R}(\mathbf{r})+\frac{\hbar \Omega _{1}g_{c}}{\Delta _{a}}\cos (kx)(\hat{a%
}+\hat{a}^{\dag })  \notag \\
&&+\frac{\hbar \Omega _{2}g_{c}}{\Delta _{a}}\cos (kx+\vartheta )(\hat{a}%
e^{i\theta }+\hat{a}^{\dag }e^{-i\theta })  \notag \\
&&+\frac{\hbar \Omega _{1}^{2}}{\Delta _{a}}\cos ^{2}(kx)+\frac{\hbar \Omega
_{1}^{2}}{\Delta _{a}}\cos ^{2}(kx+\vartheta ).  \label{SH2}
\end{eqnarray}%
where $\delta _{c}=\Delta _{c}-g_{c}^{2}/\Delta _{a}$. Note that in writing
Hamiltonian~(\ref{SH2}), a gauge with $\vartheta _{2}=\vartheta $, $\theta
_{2}=\theta $, and $\vartheta _{1}=\theta _{1}=0$ has been chosen. To
describe the dynamics of $N$ atoms, we extend the single particle
Hamiltonian~(\ref{SH2}) to the second-quantized form, i.e.,%
\begin{eqnarray}
\mathcal{\hat{H}} &=&-\hbar \delta _{c}\hat{a}^{\dag }\hat{a}+\int d^{3}%
\mathbf{r\hat{\Psi}}^{\dag }\mathbf{(r)}\left[ \frac{\mathbf{\hat{p}}^{2}}{2m%
}+\hat{V}_{R}(\mathbf{r})+\frac{\hbar \Omega _{1}g_{c}}{\Delta _{a}}\right.
\notag \\
&&\times \cos (kx)(\hat{a}+\hat{a}^{\dag })+\frac{\hbar \Omega _{2}g_{c}}{%
\Delta _{a}}\cos (kx+\vartheta )  \notag \\
&&\times (\hat{a}e^{i\theta }+\hat{a}^{\dag }e^{-i\theta })+\frac{\hbar
\Omega _{1}^{2}}{\Delta _{a}}\cos ^{2}(kx)  \notag \\
&&\left. +\frac{\hbar \Omega _{1}^{2}}{\Delta _{a}}\cos ^{2}(kx+\vartheta )%
\right] \mathbf{\hat{\Psi}(r),}  \label{SHM1}
\end{eqnarray}%
where $\mathbf{\hat{\Psi}(r)}$ denotes the field operator for annihilating
an atom at position $\mathbf{r}$. We further assume $V_{R}(\mathbf{r})$ is
strong enough so that the atomic motion in the transverse direction is
frozen to the ground state. This enables us to integrate out the transverse
degrees of freedom using $\mathbf{\hat{\Psi}(r)=}\sqrt{2/\pi \rho ^{2}}\hat{%
\psi}(x)\exp [-(y^{2}+z^{2})/\rho ^{2}]$, where $\rho $ is a transverse
characteristic length. The simplified one-dimensional Hamiltonian thus reads%
\begin{eqnarray}
\mathcal{\hat{H}} &=&-\hbar \delta _{c}\hat{a}^{\dag }\hat{a}+\int d^{3}x%
\hat{\psi}^{\dag }(x)\left[ -\frac{\hbar ^{2}}{2m}\frac{\partial ^{2}}{%
\partial x^{2}}\right.  \notag \\
&&+\hbar \eta _{2}\cos (kx+\vartheta )(\hat{a}e^{i\theta }+\hat{a}^{\dag
}e^{-i\theta })  \notag \\
&&+\hbar \eta _{1}\cos (kx)(\hat{a}+\hat{a}^{\dag })+\hbar V_{1}\cos ^{2}(kx)
\notag \\
&&\left. +\hbar V_{2}\cos ^{2}(kx+\vartheta )\right] \hat{\psi}(x),
\label{SHM2}
\end{eqnarray}%
where $V_{1,2}=\Omega _{1,2}^{2}/\Delta _{a}$ and $\eta _{1,2}=\Omega
_{1,2}g_{c}/\Delta _{a}$. By setting $\vartheta =\pi /2$, Eq.~(\ref{SHM2})
reduces to Hamiltonian~(\ref{HM}) in the main text.

\begin{figure*}[tp]
\centering
\includegraphics[width=17cm]{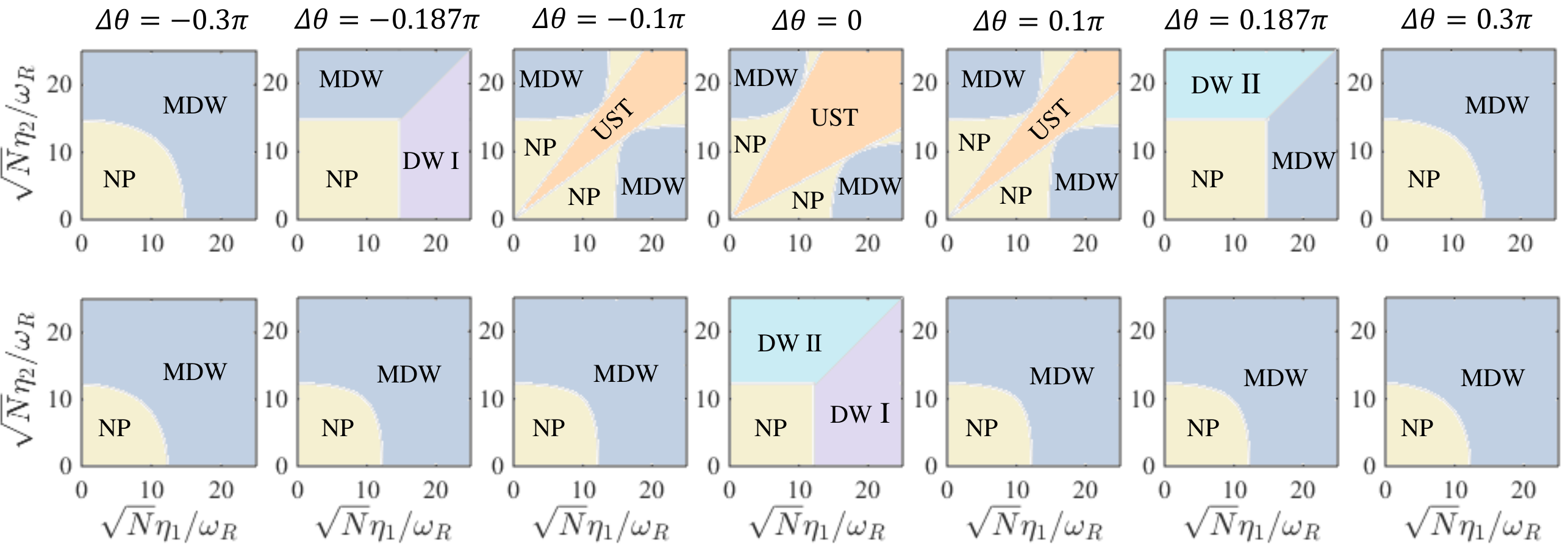} \newline
\caption{Phase diagrams with $\protect\delta _{c}/\protect\omega _{R}=-300$
and different $\Delta \protect\theta $, where $\Delta \protect\theta =%
\protect\theta -\protect\pi /2$. The top and bottom panels correspond to $%
\protect\kappa /\protect\omega _{R}=200$ and $\protect\kappa /\protect\omega %
_{R}=0$, respectively.}
\label{PhaseSI}
\end{figure*}

\section{Mean-field equations}

The Heisenberg equations of the photon annihilation operator $\hat{a}$ and
the matter wave field operator $\hat{\psi}(x)$ is derived by using the
Hamiltonian $\mathcal{\hat{H}}$,
\begin{equation}
i\frac{\partial }{\partial t}\hat{a}=\frac{1}{\hbar }[\hat{a},\mathcal{\hat{H%
}}]=(-\hbar \delta _{c}-i\hbar \kappa )\hat{a}+\eta _{1}\hat{\Theta}%
_{1}+\eta _{2}e^{-i\theta }\hat{\Theta}_{2},  \label{EoM1A}
\end{equation}%
\begin{equation}
i\frac{\partial }{\partial t}\hat{\psi}(x)=\frac{1}{\hbar }[\hat{\psi}(x),%
\mathcal{\hat{H}}]=\frac{1}{\hbar }\hat{H}_{a}\hat{\psi}(x),  \label{EoM2P}
\end{equation}%
where $\hat{\Theta}_{1}=\int d^{3}x\hat{\psi}^{\dag }(x)\cos (kx)\hat{\psi}%
(x)$ and $\hat{\Theta}_{2}=\int d^{3}x\hat{\psi}^{\dag }(x)\sin (kx)\hat{\psi%
}(x)$. Note that we have added the cavity decay rate $\kappa $ in Eq.~(\ref%
{EoM1A}). Replacing the quantum field operators $\hat{a}$ and $\hat{\psi}(x)$
by their averages $\left\langle \hat{a}(t)\right\rangle =$ $\alpha
(t)=\left\vert \alpha (t)\right\vert e^{i\phi (t)}$ and $\left\langle \hat{%
\psi}(x,t)\right\rangle =\sqrt{N}\psi (x,t)=\sqrt{Nn(x,t)}e^{i\tau }$,
respectively, we get the mean-field equations~(\ref{EoM1})-(\ref{EoM2}) in
the main text.

\section{More phase diagrams}

As plotted in Fig.~\ref{PhaseSI}, we provide more phase diagrams to show the
contrast between the dissipative (top panel) and dissipationless (bottom
panel) systems.

\section{Diagrams of the order parameters}

\begin{figure*}[tp]
\centering
\includegraphics[width=17cm]{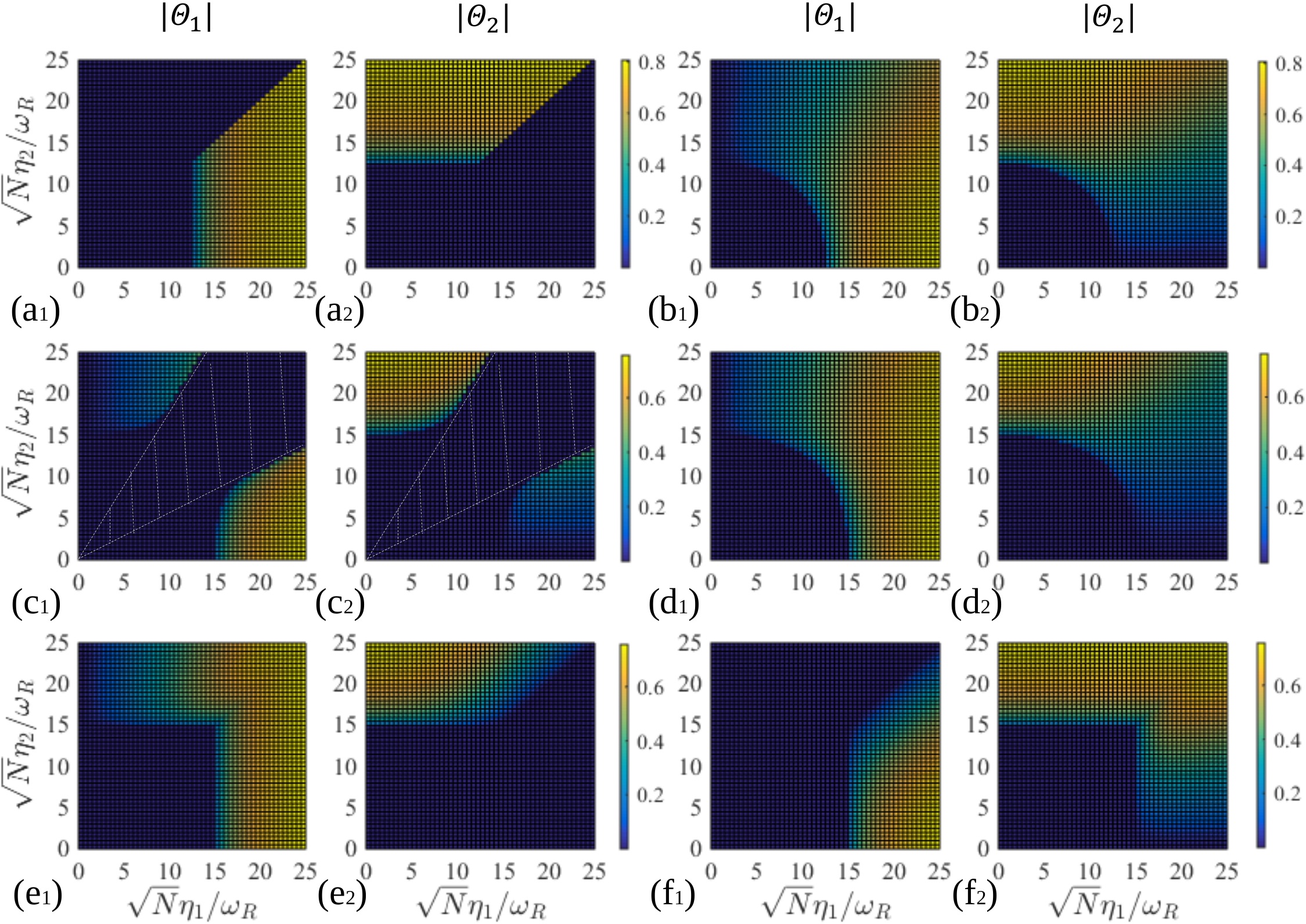} \newline
\caption{Order parameters $\left\vert \Theta _{1}\right\vert $ and $%
\left\vert \Theta _{2}\right\vert $ for (a1)-(a2) $\protect\kappa /\protect%
\omega _{R}=0$, $\protect\theta =\protect\pi /2$, (b1)-(b2) $\protect\kappa /%
\protect\omega _{R}=0$, $\protect\theta =\protect\pi /5$, (c1)-(c2) $\protect%
\kappa /\protect\omega _{R}=200$, $\protect\theta =\protect\pi /2$,
(d1)-(d2) $\protect\kappa /\protect\omega _{R}=200$, $\protect\theta =%
\protect\pi /5$, (e1)-(e2) $\protect\kappa /\protect\omega _{R}=200$, $%
\protect\theta =4.12$, and (f1)-(f2) $\protect\kappa /\protect\omega %
_{R}=200 $, $\protect\theta =-0.98$, with $\protect\delta _{c}/\protect%
\omega _{R}=-300$. The shaded areas in (c1) and (c2) indicate the absence of
stationary steady-state solutions.}
\label{PhaseSII}
\end{figure*}

Figure~\ref{PhaseSII} shows the steady-state solutions of order parameters $%
\Theta _{1}$ and $\Theta _{2}$ with the same parameters as those in Figs.~%
\ref{PhaseI} and \ref{PhaseII}, obtained by numerically solving Eqs.~(\ref%
{EoM1})-(\ref{EoM2}). In these phase diagrams, Figs.~\ref{PhaseSII}(a$i$)-%
\ref{PhaseSII}(b$i$) correspond to Figs.~\ref{PhaseI}(a) and \ref{PhaseI}(b)
and Figs.~\ref{PhaseSII}(c$i$)-\ref{PhaseSII}(f$i$) correspond to Figs.~\ref%
{PhaseII}(a)-\ref{PhaseII}(d) with $i\in \{1,2\}$, respectively. It should
be noticed that, within the shaded area in Figs.~\ref{PhaseSII}(c1) and \ref%
{PhaseSII}(c2), the system loses stationary steady-state solutions but
features limit-cycle oscillations in the long-time limit.

\section{Stability analysis beyond adiabatic elimination}

We go beyond adiabatic elimination by incorporating the dynamics of the
cavity fluctuations $\delta \alpha $ and $\delta \alpha ^{\ast }$. Assuming $%
\psi (x,t)=e^{-i\mu t/\hbar }[\psi _{0}(x)+\delta \psi (x,t)]$ and $\alpha
(t)=\alpha _{0}+\delta \alpha $, where $\psi _{0}(x)$ and $\alpha _{0}$ are
the steady-state solution of Eqs.~(\ref{EoM1})-(\ref{EoM2}) in the main
text. The equations of motion linearized in $\delta \psi $ and $\delta
\alpha $ read%
\begin{eqnarray}
i\hbar \frac{\partial }{\partial t}\delta \psi _{-} &&\!\!\!=\!\left( -\frac{%
\hbar ^{2}}{2m}\frac{\partial ^{2}}{\partial x^{2}}-\mu \right) \delta \psi
\notag \\
&&+\psi _{0}\eta _{1}\cos (kx)(\delta \alpha +\delta \alpha ^{\ast })  \notag
\\
&&+\psi _{0}\eta _{2}\sin (kx)(\delta \alpha e^{i\theta }+\delta \alpha
^{\ast }e^{-i\theta }),  \label{SFEoM1}
\end{eqnarray}%
\begin{eqnarray}
i\hbar \frac{\partial }{\partial t}\delta \alpha &\!\!=\!\!&N\eta _{1}\int
dx\cos (kx)(\psi _{0}^{\ast }\delta \psi +\psi _{0}\delta \psi ^{\ast })
\notag \\
&&+N\psi _{0}\eta _{2}\int dx\cos (kx)(\psi _{0}^{\ast }\delta \psi +\psi
_{0}\delta \psi ^{\ast })e^{-i\theta }  \notag \\
&&(-\hbar \delta _{c}-i\hbar \kappa )\delta \alpha .  \label{SFEoM2}
\end{eqnarray}%
Following the strategy employed in Sec.~\ref{sec:stability}, we substitute
the ansate $\delta \psi (x,t)=\delta \psi _{+}(x)e^{-i\omega t/\hbar
}+\delta \psi _{-}^{\ast }(x)e^{-i\omega ^{\ast }t/\hbar }$ and $\delta
\alpha (t)=\delta \alpha _{+}e^{-i\omega t/\hbar }+\delta \alpha _{-}^{\ast
}e^{i\omega ^{\ast }t/\hbar }$ into Eqs.~(\ref{SFEoM1})-(\ref{SFEoM2}) and
obtain%
\begin{eqnarray}
\hbar \omega \delta \psi _{+} &&\!\!\!=\!\left( -\frac{\hbar ^{2}}{2m}\frac{%
\partial ^{2}}{\partial x^{2}}-\mu \right) \delta \psi _{+}  \notag \\
&&+\psi _{0}\eta _{1}\cos (kx)(\delta \alpha _{+}+\delta \alpha _{-})  \notag
\\
&&+\psi _{0}\eta _{2}\sin (kx)(\delta \alpha _{+}e^{i\theta }+\delta \alpha
_{-}e^{-i\theta }),
\end{eqnarray}%
\begin{eqnarray}
\hbar \omega \delta \psi _{-} &&\!\!\!=\!\left( \frac{\hbar ^{2}}{2m}\frac{%
\partial ^{2}}{\partial x^{2}}+\mu \right) \delta \psi _{-}  \notag \\
&&-\psi _{0}\eta _{1}\cos (kx)(\delta \alpha _{+}+\delta \alpha _{-})  \notag
\\
&&-\psi _{0}\eta _{2}\sin (kx)(\delta \alpha _{+}e^{i\theta }+\delta \alpha
_{-}e^{-i\theta }),
\end{eqnarray}%
\begin{eqnarray}
\hbar \omega \delta \alpha _{+} &\!\!\!=\!\!\!&N\eta _{1}\int dx\cos
(kx)(\psi _{0}^{\ast }\delta \psi _{+}+\psi _{0}\delta \psi _{-})  \notag \\
&&+N\eta _{2}\int dx\sin (kx)(\psi _{0}^{\ast }\delta \psi _{+}+\psi
_{0}\delta \psi _{-})e^{-i\theta }  \notag \\
&&+(-\hbar \delta _{c}+i\hbar \kappa )\delta \alpha _{+},
\end{eqnarray}%
\begin{eqnarray}
\hbar \omega \delta \alpha _{-} &\!\!\!=\!\!\!&-N\psi _{0}\eta _{1}\int
dx\cos (kx)(\psi _{0}^{\ast }\delta \psi _{+}+\psi _{0}\delta \psi _{-})
\notag \\
&&-N\eta _{2}\int dx\sin (kx)(\psi _{0}^{\ast }\delta \psi _{+}+\psi
_{0}\delta \psi _{-})e^{-i\theta }  \notag \\
&&+(\hbar \delta _{c}-i\hbar \kappa )\delta \alpha _{-}.
\end{eqnarray}%
These equations can be recast in a matrix form $\omega \mathbf{f}=\mathcal{M}%
\mathbf{f}$, with $\mathbf{f=}(\delta \psi _{+},\delta \psi _{-},\delta
\alpha _{+},\delta \alpha _{-})^{\mathbf{T}}$, and
\begin{widetext}
\begin{equation}
\mathcal{M}=\left(
\begin{array}{cccc}
H_{k}-\mu  & 0 & \psi _{0}(K_{1}(x)+K_{2}(x)e^{i\theta }) & \psi
_{0}(K_{1}(x)+K_{2}(x)e^{-i\theta }) \\
0 & -H_{k}+\mu  & -\psi _{0}(K_{1}(x)+K_{2}(x)e^{i\theta }) & -\psi
_{0}(K_{1}(x)+K_{2}(x)e^{-i\theta }) \\
N(\eta _{1}\mathcal{I}_{+\ast }+\eta _{2}e^{-i\theta }\mathcal{I}_{-\ast })
& N(\eta _{1}\mathcal{I}_{+}+\eta _{2}e^{-i\theta }\mathcal{I}_{-}) &
-\delta _{c}+i\kappa  & 0 \\
-N(\eta _{1}\mathcal{I}_{+\ast }+\eta _{2}e^{-i\theta }\mathcal{I}_{-\ast })
& -N(\eta _{1}\mathcal{I}_{+}+\eta _{2}e^{-i\theta }\mathcal{I}_{-}) & 0 &
\delta _{c}-i\kappa
\end{array}%
\right) ,
\end{equation}%
\end{widetext}where $K_{1}(x)=\eta _{1}\cos (kx)$, $K_{2}(x)=\eta _{2}\sin
(kx)$ and $H_{k}=-(\hbar ^{2}/2m)\partial _{x}^{2}$ is the kinetic energy.

Using the trivial solution ($\psi _{0}(x)=1/\sqrt{L}$, $\alpha _{0}=0$), and
the ansates $\delta \psi _{\pm }=\delta \psi _{\pm }^{(1)}\cos (kx)+\delta
\psi _{\pm }^{(2)}\sin (kx)$, the dynamical matrix takes the following $%
6\times 6$ form,%
\begin{widetext}
\begin{equation}
\mathcal{\tilde{M}=}\left(
\begin{array}{cccccc}
\omega _{R} & 0 & 0 & 0 & \eta _{1} & \eta _{1} \\
0 & -\omega _{R} & 0 & 0 & -\eta _{1} & -\eta _{1} \\
0 & 0 & \omega _{R} & 0 & \eta _{2}e^{i\theta } & \eta _{2}e^{-i\theta } \\
0 & 0 & 0 & -\omega _{R} & -\eta _{2}e^{i\theta } & -\eta _{2}e^{-i\theta }
\\
N\eta _{1}/2 & N\eta _{1}/2 & N\eta _{2}e^{-i\theta }/2 & N\eta
_{2}e^{-i\theta }/2 & -\delta _{c}+i\kappa  & 0 \\
-N\eta _{1}/2 & -N\eta _{1}/2 & -N\eta _{2}e^{i\theta }/2 & -N\eta
_{2}e^{i\theta }/2 & 0 & \delta _{c}-i\kappa
\end{array}%
\right) .
\end{equation}%
\end{widetext}

The eigenvalues $\omega $ of $\mathcal{\tilde{M}}$ are the solutions of the
sixth-order characteristic equation Det$(\mathcal{M}-\omega I_{6\times 6})=0$%
, namely the solutions of 
\begin{eqnarray}
&&\left[ (\delta _{c}\omega _{R}+2N\eta _{1}^{2})\omega _{R}+\delta
_{c}\omega ^{2}\right] \left[ (\delta _{c}\omega _{R}+2N\eta _{2}^{2})\omega
_{R}+\delta _{c}\omega ^{2}\right]  \notag \\
&=&(\omega +i\kappa )^{2}(\omega _{R}^{2}-\omega ^{2})^{2}+4\omega
_{R}^{2}N^{2}\eta _{1}^{2}\eta _{2}^{2}\cos ^{2}(\theta ).
\end{eqnarray}

\section{Steady-state quantum phases for the effective three-level model}

In this section, we describe the methods in obtaining the phase diagram of
the effective three-level model in more detail. Choosing the state $%
\left\vert 0\right\rangle $ as a reference, we apply the generalized
Holstein-Primakoff transformation \cite{PRAB13,PRACC13} to rewrite the
operators $\hat{\Xi}_{ij}$ as
\begin{eqnarray}
\hat{\Xi}_{00} &=&N-\sum_{i=1,2}b_{i}^{\dag }b_{i}\text{,} \\
\text{ }\hat{\Xi}_{12} &=&b_{1}^{\dag }b_{2}\text{, } \\
\hat{\Xi}_{s0} &=&b_{s}^{\dag }\sqrt{N-\sum_{i=1,2}b_{i}^{\dag }b_{i}}\text{
}(s=1,2)\text{,}
\end{eqnarray}%
where $b_{i}^{\dag }$ and $b_{i}$ are bosonic operators. In order to
construct a mean-field theory, the bosonic operators are assumed to be
composed of their expectation value and a fluctuation operator, i.e.,
\begin{equation}
a=\alpha +\delta a\text{, }b_{1}=\beta _{1}+\delta b_{1}\text{, }b_{2}=\beta
_{2}+\delta b_{2},  \label{SOP}
\end{equation}%
where $\alpha =\left\langle a\right\rangle $, $\beta _{1}=\left\langle
b_{1}\right\rangle $, and $\beta _{2}=\left\langle b_{2}\right\rangle $ are
complex mean-field parameters. According to Eq.~(\ref{SOP}), the operators $%
\hat{\Xi}_{ij}$ can be expanded as%
\begin{eqnarray*}
\hat{\Xi}_{00} &=&Np-\beta _{1}\delta b_{1}^{\dag }-\beta _{2}\delta
b_{2}^{\dag }-\beta _{1}^{\ast }\delta b_{1}-\beta _{2}^{\ast }\delta b_{2}
\\
&&+\mathcal{O}(\delta b_{1,2})^{2}, \\
\hat{\Xi}_{12} &=&\beta _{1}^{\ast }\beta _{2}+\beta _{2}\delta b_{1}^{\dag
}+\beta _{1}\delta b_{2}^{\dag }+\mathcal{O}(\delta b_{1,2})^{2}, \\
\hat{\Xi}_{i0} &=&\sqrt{Np}\delta b_{i}^{\dag }+\sqrt{p}\beta _{2}^{\ast }+%
\mathcal{O}(\delta b_{1,2})^{2}\text{ \ \ \ \ \ (}i=1,2\text{),} \\
\hat{\Xi}_{ii} &=&\left\vert \beta _{i}\right\vert ^{2}+\beta _{i}\delta
b_{i}^{\dag }+\beta _{i}^{\ast }\delta b_{i}+\mathcal{O}(\delta b_{1,2})^{2}%
\text{ \ \ (}i=1,2\text{),}
\end{eqnarray*}%
where $p=\sqrt{1-\left\vert \beta _{1}\right\vert ^{2}-\left\vert \beta
_{2}\right\vert ^{2}}$.\ In terms of the mean-field parameters $\alpha $ and
$\beta _{i}$ ($i=1,2$), the semi-classical equations of motion, $i\hbar
\partial _{t}\left\langle \hat{\Xi}_{ij}\right\rangle =\left\langle [\hat{\Xi%
}_{ij},\mathcal{\hat{H}}]\right\rangle $ and $i\hbar \partial
_{t}\left\langle \hat{a}\right\rangle =\left\langle [\hat{a},\mathcal{\hat{H}%
}]\right\rangle -i\hbar \kappa \left\langle \hat{a}\right\rangle $, are
derived as%
\begin{eqnarray}
i\frac{\partial }{\partial t}\beta _{1} &=&-\omega _{R}\beta _{1}-\frac{\mu
_{1}(\alpha +\alpha ^{\ast })(\left\vert \beta _{1}\right\vert ^{2}-N)}{%
\sqrt{p}}  \notag \\
&&-\frac{\mu _{2}(\alpha e^{i\theta }+\alpha ^{\ast }e^{-i\theta })\beta
_{2}^{\ast }\beta _{1}}{\sqrt{p}},  \label{SEoM1}
\end{eqnarray}%
\begin{eqnarray}
i\frac{\partial }{\partial t}\beta _{2} &=&-\omega _{R}\beta _{2}-\frac{\mu
_{2}(\alpha +\alpha ^{\ast })(\left\vert \beta _{2}\right\vert ^{2}-N)}{%
\sqrt{p}}  \notag \\
&&-\frac{\mu _{1}(\alpha e^{i\theta }+\alpha ^{\ast }e^{-i\theta })\beta
_{1}^{\ast }\beta _{2}}{\sqrt{p}},  \label{SEoM2}
\end{eqnarray}%
\begin{eqnarray}
i\frac{\partial }{\partial t}\alpha  &=&(-\delta _{c}-i\kappa )\alpha +\mu
_{1}\sqrt{p}(\beta _{1}+\beta _{1}^{\ast })  \notag \\
&&+\mu _{2}\sqrt{p}e^{-i\theta }(\beta _{2}+\beta _{2}^{\ast }).
\label{SEoM3}
\end{eqnarray}%
Following the same manner we did in Sec.~\ref{sec:stability} of the main
text, the stability of the steady-state solutions of Eqs.~(\ref{SEoM1})-(\ref%
{SEoM3}) are determined by analyzing the linearized fluctuation equations, $i%
\mathbf{\dot{f}}_{\text{T}}=\mathcal{M}_{\text{T}}\mathbf{f}_{\text{T}}$,
with $\mathbf{f}_{\text{T}}\mathbf{=}(\delta \psi _{+},\delta \psi
_{-},\delta \alpha _{+},\delta \alpha _{-})^{\mathbf{T}}$ and
\begin{widetext}
\begin{equation}
\mathcal{M}_{\text{T}}=\left(
\begin{array}{cccccc}
-\Delta -i\kappa  & 0 & \mu _{1}\sqrt{p} & \mu _{1}\sqrt{p} & \mu _{2}\sqrt{p%
}e^{-i\theta } & \mu _{2}\sqrt{p}e^{-i\theta } \\
0 & \Delta +i\kappa  & -\mu _{1}\sqrt{p} & -\mu _{1}\sqrt{p} & -\mu _{2}%
\sqrt{p}e^{i\theta } & -\mu _{2}\sqrt{p}e^{i\theta } \\
-B_{1}^{\ast }(-\theta ) & -B_{1}^{\ast }(\theta ) & \omega _{R}-\Lambda
_{1}^{\ast } & -2\mu _{1}\beta _{1}^{\ast }\varrho (0) & -\mu _{1}\beta
_{2}^{\ast }\varrho (0)-\mu _{2}\beta _{1}^{\ast }\varrho (\theta ) & -\mu
_{1}\beta _{2}^{\ast }\varrho (0) \\
B_{1}(\theta ) & B_{1}(-\theta ) & 2\mu _{1}\beta _{1}\varrho (0) & -\omega
_{R}+\Lambda _{1} & \mu _{1}\beta _{2}\varrho (0)+\mu _{2}\beta _{1}\varrho
(\theta ) & \mu _{1}\beta _{2}\varrho (0) \\
-B_{2}^{\ast }(-\theta ) & -B_{2}^{\ast }(\theta ) & -\mu _{2}\beta
_{1}^{\ast }\varrho (\theta ) & -\mu _{1}\beta _{2}\varrho (0)-\mu _{2}\beta
_{1}^{\ast }\varrho (\theta ) & \omega _{R}-\Lambda _{2}^{\ast } & -2\mu
_{2}\beta _{1}^{\ast }\varrho (\theta ) \\
B_{2}(\theta ) & B_{2}(-\theta ) & \mu _{1}\beta _{2}^{\ast }\varrho (0)+\mu
_{2}\beta _{1}\varrho (\theta ) & \mu _{2}\beta _{1}\varrho (\theta ) & 2\mu
_{2}\beta _{1}\varrho (\theta ) & -\omega _{R}+\Lambda _{2}%
\end{array}%
\right) .  \label{SFEoM}
\end{equation}%
\end{widetext}
Here $\varrho (\theta )=(\alpha \exp (i\theta )+\alpha ^{\ast }\exp
(-i\theta ))/\sqrt{p}$, $B_{1}(\theta )=[\mu _{1}(\left\vert \beta
_{1}\right\vert ^{2}-p)+\mu _{2}\beta _{1}^{\ast }\beta _{2}\exp (i\theta )]/%
\sqrt{p}$, $B_{2}(\theta )=[\mu _{2}(\left\vert \beta _{2}\right\vert
^{2}-p)+\mu _{1}\beta _{2}^{\ast }\beta _{1}\exp (i\theta )]/\sqrt{p}$, $%
\Lambda _{1}=2\mu _{1}\beta _{1}\varrho (0)+\mu _{2}\beta _{2}\varrho
(\theta )$, and $\Lambda _{2}=\mu _{1}\beta _{1}\varrho (0)+2\mu _{2}\beta
_{2}\varrho (\theta )$. From Eqs.~(\ref{SEoM1})-(\ref{SFEoM}), the
mean-field parameters characterizing different quantum phases can be
uniquely determined.

The solutions in the case of $\theta =\pi /2$\ and $\kappa =0$ are
summarized as follows. Firstly, for $(\mu _{1},\mu _{2})<\mu _{c}$, with $%
\mu _{c}=\sqrt{-\delta _{c}\omega _{R}}/2\equiv \eta _{c}\sqrt{2N}/2$, both $%
\left\langle \hat{\Xi}_{11}\right\rangle $ and $\left\langle \hat{\Xi}%
_{22}\right\rangle $ vanish, which defines the NP. Secondly,\ for $\mu
_{1}>\mu _{c}$ and $\mu _{1}>\mu _{2}$, we have $\left\langle \hat{\Xi}%
_{11}\right\rangle /N=(4\mu _{1}^{2}+\delta _{c}\omega _{R})/8\mu _{1}^{2}$
and $\left\langle \hat{\Xi}_{22}\right\rangle /N=0$. This means that the
atoms start populating the state $\left\vert 1\right\rangle $, which
corresponds to the DW \textrm{I}. Thirdly, for $\mu _{2}>\mu _{c}$ and $\mu
_{2}>\mu _{1}$, we obtain $\left\langle \hat{\Xi}_{22}\right\rangle /N=(4\mu
_{2}^{2}+\delta _{c}\omega _{R})/8\mu _{2}^{2}$ and $\left\langle \hat{\Xi}%
_{11}\right\rangle /N=0$, indicating the state $\left\vert 2\right\rangle $
is occupied. This corresponds to the DW \textrm{II}. Lastly, for $\mu
_{1}=\mu _{2}>\mu _{c}$, the values of $\left\langle \hat{\Xi}%
_{11}\right\rangle $ and $\left\langle \hat{\Xi}_{22}\right\rangle $ are
determined by the equation $\left\langle \hat{\Xi}_{11}\right\rangle
/N+\left\langle \hat{\Xi}_{22}\right\rangle /N=(4\mu _{1,2}^{2}+\delta
_{c}\omega _{R})/8\mu _{1,2}^{2}$, signaling both $\left\vert 1\right\rangle
$ and $\left\vert 2\right\rangle $ can be populated, and thus the MDW is
realized.

Notice that analytical solutions for more generic parameters are not
available. However, it can still be straightforwardly found that the
mean-field parameters satisfying $\beta _{1}\beta _{2}=0$ and $\beta
_{1}+\beta _{2}\neq 0$ could by no means be a steady-state solution of Eqs.~(%
\ref{SEoM1})-(\ref{SEoM3}), except for the case of $\theta =\pi /2$\ and $%
\kappa =0$. This implies that, at least under the framework of three-mode
approximation, the DW \textrm{I }and DW \textrm{II }can not exist\textrm{\ }%
for any other parameter settings.

\end{document}